\documentclass[aps,pre,twocolumn,showpacs]{revtex4}
\usepackage{amsmath}
\usepackage{amsfonts}
\usepackage{amssymb}
\usepackage{epsfig}

\newcommand{\be}{\begin{equation}}
\newcommand{\ee}{\end{equation}}
\newcommand{\ra}{\rangle}
\newcommand{\la}{\langle}

\begin{document}

\title{Large Fluctuations of the Macroscopic Current in Diffusive Systems: \\ A Confirmation of the Additivity Principle}

\author{Pablo I. Hurtado}
\author{Pedro L. Garrido}
\affiliation{Departamento de Electromagnetismo y F\'{\i}sica de la Materia, 
and Instituto Carlos I \\
de F\'{\i}sica Te\'orica y Computacional, Universidad de Granada, 
Granada 18071, Spain}

\date{\today}

\begin{abstract}
Most systems, when pushed out of equilibrium, respond by building up currents of locally-conserved observables.
Understanding how microscopic dynamics determines the averages and fluctuations of these currents is one of the main open problems in nonequilibrium statistical physics.
The additivity principle is a theoretical proposal that allows to compute the current distribution in many one-dimensional nonequilibrium systems. Using simulations, 
we confirm this conjecture in a simple and general model of energy transport, both in the presence of a temperature gradient and in canonical equilibrium. In particular, we show that the current distribution 
displays a Gaussian regime for small current fluctuations, as prescribed by the central limit theorem, and non-Gaussian (exponential) tails for large current deviations,  
obeying in all cases the Gallavotti-Cohen fluctuation theorem. In order to facilitate a given current fluctuation, the system adopts
a well-defined temperature profile different from that of the steady state, and in accordance with the additivity hypothesis predictions.
System statistics during a large current fluctuation is independent of the sign of the current, which implies that 
the optimal profile (as well as higher-order profiles and spatial correlations) are invariant upon currenst inversion.
We also demonstrate that finite-time joint fluctuations of the current and the profile are well described by the additivity functional.
These results confirm the additivity hypothesis as a general and powerful tool to compute current distributions in many nonequilibrium systems. 
\end{abstract}

\maketitle

\section{Introduction}
\label{intro}
Understanding the physics of systems out of equilibrium remains challenging to a large extent, even in the simplest setting for which one could expect to make 
significant advances, which is that of a nonequilibrium steady state (NESS).
Even in this simple situation difficulties abound mainly because out of equilibrium 
the dynamics plays a dominant role \cite{Spohn,Marro}.
For instance, the phase space available to a system in a NESS depends crucially on the dynamics, resulting in a probability measure for microscopic configurations
which is not known in general for a NESS, as it will inherit this dependence on the dynamics \cite{SRB}. This is in contrast to the equilibrium case, where the available phase space is uniquely 
determined by the Hamiltonian and the Gibss distribution provides the probability measure for microscopic configurations.
One can ask however questions on the statistics of the macroscopic observables
characterizing a NESS, as for instance the current flowing through the system \cite{Bertini,BD,Derrida,Derrida2}.
In equilibrium, the fluctuations of macroscopic quantities, which are a reflection of the hectic microscopic world, are strikingly independent
of microscopic details, being solely determined by thermodynamic quantities as
the entropy, free energy, etc. A natural way to seek a macroscopic theory of nonequilibrium 
phenomena is thus to investigate the fluctuations of macroscopic currents.
Unveiling the relation between microscopic dynamics and current fluctuations
has proven to be a 
difficult task \cite{Bertini,BD,Derrida,Derrida2,Derrida3,Livi,Dhar,we,GC,LS}, and up to now only few exactly-solvable cases are understood.
An important step in this direction has been the development of the Gallavotti-Cohen fluctuation theorem \cite{GC,LS}, 
which relates the probability of forward and backward currents reflecting the time-reversal symmetry of microscopic dynamics. 
However, we still lack a general approach based on few simple principles.
Recently, Bertini, De Sole, Gabrielli, Jona-Lasinio and Landim \cite{Bertini} have introduced a Hydrodynamic Fluctuation Theory (HFT) 
to study large dynamic fluctuations of diffusive systems. This is a very general approach which leads to a hard
variational problem whose solution remains challenging in most cases.
Simultaneously, Bodineau and Derrida \cite{BD,Derrida,Derrida2} have conjectured an additivity principle for current fluctuations in one dimension 
which can be readily applied to obtain quantitative predictions and, together with HFT,
seems to open the door to a general theory for nonequilibrium systems.

In this paper we test in depth the validity of the additivity principle in a simple and very general diffusive model. In particular, we investigate the fluctuations of 
the energy current in the one-dimensional (1D) Kipnis-Marchioro-Pressuti (KMP) model of heat conduction, which
represents at a coarse-grained level a large class of quasi-1D diffusive systems of technological and theoretical interest for which 
understanding current statistics is of central importance. Our results strongly support the validity of the additivity principle to describe current fluctuations
in one dimension, both in the presence of a temperature gradient (NESS) and in canonical equilibrium. 
In particular, we find that the current distribution shows both Gaussian and non-Gaussian regimes, and obeys the Gallavotti-Cohen symmetry. 
The system modifies its temperature profile to facilitate a given current fluctuation, as predicted by the theory, and this profile (as well as any other higher-order profile
and spatial correlation) turns out to be independent of the sign of the current. 
We also explore physics beyond the additivity conjecture by studying the fluctuations of the total energy in the system, which exhibit the trace left by corrections
to local equilibrium resulting from the presence of weak long-range correlations in the NESS. In addition, we extend the additivity hypothesis to study the joint fluctuations of the
current and the profile. 

The paper is structured as follows. In next section we describe the additivity principle from a general perspective. Section \ref{model} introduces 
the KMP model in one dimension. In section \ref{numres} we report the results of our simulations, together with a detailed comparison with theoretical predictions. 
Here we also show evidence of structure beyond the additivity scenario. 
Section \ref{joint} investigates the joint fluctuations of the current and the 
temperature profile, extending the additivity principle to understand these finite-time corrections. Finally, we present 
our conclusions in section \ref{conclu}, and a number of appendices describe some technical aspects of the discussion in the main text.
Part of the work reported in this paper was presented in a shorter Letter \cite{Pablo}.

\section{The Additivity Principle}
\label{addit}

The additivity principle (to which we will also refer here as BD theory) is a conjecture first proposed by T. Bodineau and B. Derrida \cite{BD} 
that enables one to calculate the fluctuations of the current in 1D diffusive 
systems in contact with two boundary thermal baths at different temperatures, $T_L\neq T_R$.
It is a very general conjecture of broad applicability, expected to hold for 1D systems of classical interacting particles,
both deterministic or stochastic, independently of the details of the interactions between the particles or the coupling to the thermal reservoirs. 
The only requirement is that the system at hand must be diffusive, i.e. Fourier's law must hold. If this is the case, the additivity principle predicts 
the full current distribution in terms of its first two cumulants.
Equivalently, one may use the same formalism to study 
diffusive particle systems coupled to particle reservoirs at the boundaries at different chemical potentials, and 
obeying Fick's law, or any other open diffusive system characterized by a single locally-conserved field. 
However, in this paper we stick for simplicity to the energy-diffusion version of the problem. 
Let $\text{P}_N(q,T_L,T_R,t)$ be the probability of observing a time-integrated current $Q_t=q t$ during a long 
time $t$ in a system of size $N$. This probability typically obeys a large deviation principle \cite{LD,LDH},
\begin{equation}
\text{P}_{N}(q,T_L,T_R,t)\sim \text{e}^{+t {\cal F}_{N}(q,T_L,T_R)} \, ,
\label{pn}
\end{equation}
where ${\cal F}_{N}(q,T_L,T_R)$ is the current large-deviation function (LDF), such that ${\cal F}_{N}(\langle q \rangle,T_L,T_R)=0$ and 
${\cal F}_{N}(q\ne\langle q \rangle,T_L,T_R)<0$, with $\la q\ra= \lim_{t\to \infty} Q_t/t$. This means in particular that current fluctuations away from the average are exponentially 
unlikely in time. The additivity principle relates this probability with the product of probabilities for sustaining the same current in subsystems of lengths $N-n$ and $n$, 
\begin{equation}
{\text{P}_{N}(q,T_L,T_R,t) = \max_T \left[ \text{P}_{N-n}(q,T_L,T,t)\, \text{P}_{n}(q,T,T_R,t) \right] } \, . 
\label{addprob}
\end{equation}
The maximization over the contact temperature $T$ can be rationalized by writing the above probability as an integral over $T$ of the product of probabilities for 
subsystems and noticing that these should obey also a large deviation principle akin to eq. (\ref{pn}). Hence a saddle-point 
calculation in the long-$t$ limit leads to (\ref{addprob}).
The additivity principle can be then rewritten for the large deviation function as
\begin{equation}
{\cal F}_{N}(q,T_L,T_R) = \max_T \left[ {\cal F}_{N-n}(q,T_L,T) + {\cal F}_{n}(q,T,T_R) \right] \, .
\label{add}
\end{equation}
We now may adopt a scaling form ${\cal F}_{N}(q,T_L,T_R) = N^{-1} {\cal G}(Nq,T_L,T_R)$ for the current LDF \cite{BD,Derrida,Derrida2}, and proceed by slicing iteratively the 
1D system of length $N$ into smaller and smaller segments. For small enough segments the temperature difference across each of them will be small, so for small currents 
$q\sim {\cal O}(N^{-1})$ each interval can be considered to be close to equilibrium and hence exhibits locally-Gaussian fluctuations around the average current (given by Fourier's law) 
at the leading order. In this way we obtain
in the continuum limit the following variational form for ${\cal G}$ \cite{BD,Derrida,Derrida2}
\begin{equation}
{\cal G}(q)=- \min_{T_q(x)}\left\{ \int_0^1 \frac{\left[q + \kappa[T_q(x)] T'_q(x) \right]^2}{2\sigma[T_q(x)]} dx \right\} \, ,
\label{ldf1}
\end{equation}
where we dropped the dependence on the baths for convenience. Here $\kappa(T)$ is the thermal conductivity characterizing Fourier's law, 
$\langle Q_t\rangle/t=-\kappa(T)\, \nabla T$, and $\sigma(T)$ measures current fluctuations in equilibrium ($T_L=T_R$), $\langle Q_t^2\rangle/t=\sigma(T)/N$.
The optimal temperature profile $T_q(x)$ derived from (\ref{ldf1}) by functional differentiation obeys
\begin{equation}
\kappa^2[T_q(x)] \left(\frac{d T_q(x)}{dx}\right)^2 = q^2\left\{1+2\sigma[T_q(x)] K(q^2) \right\}  \, ,
\label{optprof}
\end{equation}
where $K(q^2)$ is a constant which guarantees the correct boundary conditions, $T_q(0)=T_L$ and $T_q(1)=T_R$. In what follows
we assume $T_L>T_R$ without loss of generality. Equations (\ref{ldf1}) and (\ref{optprof}) completely determine the current distribution, 
which is in general non-Gaussian (except for very small current fluctuations) and obeys the Gallavotti-Cohen symmetry, 
\be
{\cal G}(-q) = {\cal G}(q) -{\cal E}\, q \, ,
\label{GCeq}
\ee
with ${\cal E}$ a constant defined by 
\cite{BD}
\begin{equation}
{\cal E}=2\int_{T_L}^{T_R} \frac{\kappa(T)}{\sigma(T)} dT \, . \nonumber
\label{constantGC}
\end{equation} 
Moreover, the optimal profile solution of eq. (\ref{optprof}) is independent of the sign of the current, $T_q(x)=T_{-q}(x)$, a rather counter-intuitive result
which, together with the Gallavotti-Cohen relation, reflects the time-reversal symmetry of microscopic dynamics \cite{GC,LS}.

In the simplest case, when $K(q^2)$ is large enough for the rhs of eq. (\ref{optprof}) not to vanish --something that happens for currents close to the average, 
the optimal profile $T_q(x)$ is monotone and we have ($T_L>T_R$)
\begin{equation}
 \frac{d T_q(x)}{dx} = -\frac{|q|}{\kappa[T(x)]} \sqrt{1+2\sigma[T(x)] K(q^2) }  \, ,
\label{optprof1}
\end{equation}
Using this expression in eq. (\ref{ldf1}) leads to
\begin{equation}
{\cal G}(q)= \int_{T_R}^{T_L}  \frac{\kappa(T)}{\sigma(T)} \left\{q - |q| \frac{1+K(q^2) \sigma(T)}{\sqrt{1+2K(q^2) \sigma(T)}}  \right\} dT \, ,
\label{ldfmon}
\end{equation}
and integrating eq. (\ref{optprof1}) above over the whole interval $x\in[0,1]$ we obtain an implicit equation for $K(q^2)$,
\begin{equation}
|q| = \int_{T_R}^{T_L}  \frac{\kappa(T) }{\sqrt{1+2K(q^2) \sigma(T)}} \, dT \, .
\label{KK}
\end{equation}

In many applications it is interesting to work with the Legendre transform of the large deviation function, 
\be
\mu(\lambda)\equiv \frac{1}{N} \max_q\left[{\cal G}(q) + \lambda q \right] \, ,
\label{mulamb}
\ee
or equivalently $\mu(\lambda)= N^{-1} [{\cal G}(q_o) + \lambda q_o]$, 
with $q_o(\lambda)$ given by $\partial_q {\cal G}(q_o) + \lambda=0$. By noticing that $\partial_q {\cal G}(q) = {\cal G}/q + K q$, it then follows 
for monotone profiles
\begin{equation}
\mu(\lambda) = - \frac{K(\lambda)}{N} \left\{ \int_{T_R}^{T_L} \frac{\kappa(T)}{\sqrt{1+2K(\lambda) \sigma(T)}} dT \right\}^2 \, ,
\label{mumon}
\end{equation}
where $K(\lambda)$ is now obtained from
\begin{equation}
\lambda = \int_{T_R}^{T_L} \left[ \frac{\text{sgn}[q_o(\lambda)]}{\sqrt{1+2 K(\lambda) \sigma(T)}} - 1 \right] dT \, ,
\label{KKlamb}
\end{equation}
and $\text{sgn}(q)=|q|/q$ is the sign function. The function $\mu(\lambda)$ can be viewed as the conjugate \emph{potential}  
to ${\cal G}(q)$, with $\lambda$ the parameter conjugate to the current $q$, a relation equivalent to the free energy being the Legendre transform 
of the internal energy in thermodynamics, with the temperature as conjugate parameter to the entropy.

When the constant $K$ is negative enough for the rhs of eq. (\ref{optprof}) to vanish at some point, the resulting optimal profile $T_q(x)$ becomes non-monotone. In this case
it can be shown \cite{BD} that the expressions for ${\cal G}(q)$ and $K(q^2)$, or their equivalent formulas in $\lambda$-space, are just the analytic
continuation of their monotone-case counterparts. 
Appendix \ref{BDpredic} shows the particular expressions for the current LDF and the associated optimal profile, both in the monotone and non-monotonous cases, 
as derived when applying this general scheme to the particular model of interest in this paper, the Kipnis-Marchioro-Presutti (KMP) model of heat conduction \cite{kmp}.

Before continuing with the description of this model, it is worth noticing that 
the additivity principle can be better understood within the context of Hydrodynamic Fluctuation Theory of Bertini et al. \cite{Bertini}, which
provides a variational principle for the most probable (possibly time-dependent) profile responsible of a given current fluctuation.
The probability of observing a particular history of the temperature profile $T(x,t)$ and the rescaled current $j(x,t)$ during a
macroscopic time is, according to HFT \cite{Bertini,tprofile},
\be
P\left(\{T(x,t),j(x,t) \} \right) \sim \exp\left(-N{\cal I}_t[T,j] \right) \,
\label{probHFT}
\ee
where the functional ${\cal I}_t$ can be written as
\be
{\cal I}_t[T,j] = \int_0^t d\tau \int_0^1 dx \frac{\left[j(x,\tau)+\kappa[T(x,\tau)] T'(x,\tau) \right]^2}{2\sigma[T(x,\tau)]} \, ,
\label{IHFT}
\ee
and where the rescaled current field is related to the temperature profile via the continuity equation $\partial_{\tau} T(x,\tau)+\partial_x j(x,\tau)=0$.
The large deviation function of the integrated current is then
\be
P\left(\frac{Q_t}{t}=\frac{q}{N} \right) \sim \exp \left[+\frac{t}{N} {\cal G}(q) \right] \, ,
\label{QHFT}
\ee
where ${\cal G}(q)$ is related to ${\cal I}_t[T,j]$ via
\be
{\cal G}(q) = \lim_{t\to \infty} \left(-\frac{1}{t} \min_{T(x,\tau) \atop j(x,\tau)}{\cal I}_t[T,j] \right) \, ,
\label{GIHFT}
\ee
with the constraint
\be
q=\frac{1}{t} \int_0^t j(x,\tau) d\tau \, ,
\label{consHFT}
\ee
and $T(x,\tau)$ and $j(x,\tau)$ coupled via the above continuity equation. Solving this time-dependent problem to obtain explicit predictions for the current LDF 
remains a challenge in most cases. 
The additivity principle, 
which on the other hand can be readily applied to obtain quantitative predictions, 
is equivalent within HFT to the hypothesis that
the optimal profiles $T(x,\tau)$ and $j(x,\tau)$ solution of the variational problem (\ref{GIHFT})-(\ref{consHFT})  are time-independent, 
in which case we recover eq. (\ref{ldf1}) for ${\cal G}(q)$.
In some special cases this approximation breaks down for for extreme 
current fluctuations \cite{Bertini,tprofile,Pablo3}, but even so the additivity hypothesis correctly predicts the current LDF 
in a very large current interval, making it very appealing.

\section{The KMP Model}
\label{model}

\begin{figure}
\centerline{\psfig{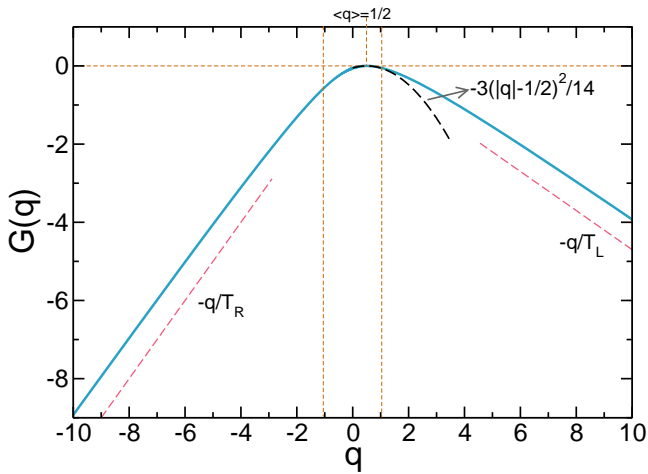}}
\caption{(Color online) ${\cal G}(q)$ for the KMP model as derived from the additivity principle, for $T_L=2$ and $T_R=1$. 
Notice the linear decay for large enough $|q|$. Vertical lines signal the crossover from monotone ($|q|<\pi/3$) 
to non-monotone ($|q|>\pi/3$) optimal profiles. The Gaussian approximation for $q \approx \langle q \rangle$, ${\cal G}(q) \approx -3(|q|-\frac{1}{2})^2/14$, is also shown.  
}
\label{predic}
\end{figure}

The system is defined on a 1D open lattice with $N$ sites \cite{kmp}. Each site models an harmonic oscillator which is mechanically uncoupled from its 
nearest neighbors but interact with them through a random process which redistributes energy locally.  In this way, a configuration is given by 
$C\equiv \{e_i, i=1\ldots N \}$, where $e_i\in \mathbb{R}_+$ is the energy of  site $i$, and the stochastic dynamics proceeds through random energy exchanges 
between randomly-chosen nearest neighbors, i.e. $(e_i,e_{i+1}) \to (e'_i,e'_{i+1})$  for $i\in[1,N-1]$ such that
\begin{eqnarray}
e'_i & = & p(e_i+e_{i+1}) \nonumber \\
e'_{i+1} & = & (1-p)(e_i+e_{i+1}) \, , 
\label{defKMP1}
\end{eqnarray}
with $p\in[0,1]$ a homogeneous random number so $e_i+e_{i+1}=e'_i+e'_{i+1}$.
In addition, boundary sites ($i=1,N$) may also exchange energy with boundary heat baths at temperatures $T_L$ for $i=1$ and $T_R$ for $i=N$,
i.e. $e_{1,N} \to e'_{1,N}$ such that
\be
e'_{1,N}=p(\tilde{e}_{L,R}+e_{1,N}) \, 
\label{defKMP2}
\ee
with $\tilde{e}_{L,R}$ randomly drawn at each step from a Gibbs distribution at the corresponding temperature, $\beta_{k}\exp (-\beta_{k} \tilde{e}_{k})$, $k=L,R$, and 
$p\in[0,1]$ random. For $T_L \neq T_R$ KMP proved \cite{kmp} that the system reaches a nonequilibrium steady state which, in the $N\to \infty$ hydrodynamic scaling limit, is described by Fourier's law with a nonzero average current
\be
\la q\ra =-\kappa(T) \frac{\text{d} T_{\text{st}}(x)}{\textrm{d} x} \quad , \quad x\in [0,1] \, ,
\label{fourier}
\ee
with $\kappa(T)=\frac{1}{2}$, and a linear energy profile
\be
T_{\text{st}}(x)=T_L + x\, (T_R - T_L) \, .
\label{stprof}
\ee
In addition, convergence to the local Gibbs measure was proven in this limit \cite{kmp}, meaning that $e_i$, $i\in[1,N]$, has an exponential distribution 
with local temperature $T_{\text{st}}[x=i/(N+1)]$ in the thermodynamic limit. However, corrections to Local Equilibrium (LE), though
vanishing in the $N\to \infty$ limit, become apparent at the fluctuation level \cite{Sp,BGL}, as we will show below.
Moreover, the fluctuations of the current in equilibrium ($T_L=T_R$) are described by $\sigma(T)=T^2$. 
It is also worth noticing that KMP dynamics obeys the local detailed balance condition and is therefore time-reversible \cite{LS}, see Appendix \ref{profil}. 
In this way we expect the Gallavotti-Cohen symmetry to hold in this system, see eq. (\ref{GCeq}).

\begin{figure}
\centerline{\psfig{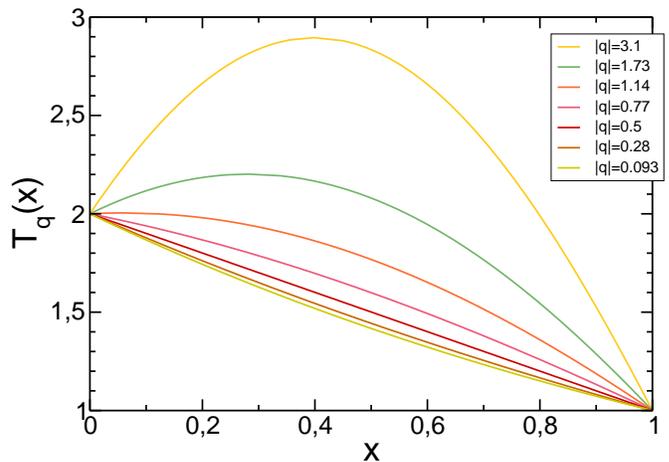}}
\caption{(Color online) Optimal $T_q(x)$ for different values of $|q|$, both in the monotone and non-monotone regimes, for $T_L=2$ and $T_R=1$.
The optimal profiles are independent of the sign of the current, $T_q(x)=T_{-q}(x)$.
}
\label{predicprof}
\end{figure}

The KMP model plays a fundamental role in nonequilibrium statistical physics as a benchmark to test new theoretical advances, and represents 
at a coarse-grained level a large class of quasi-1D diffusive systems of technological and theoretical interest. In this way, understanding how the 
energy current fluctuates in the KMP model is of central importance to understand current statistics in more realistic systems.
Furthermore, the KMP model is an optimal candidate to test the additivity principle because: (i) One can solve eqs. (\ref{ldf1}) and (\ref{optprof}) to obtain 
explicit predictions for its current LDF, and (ii) its simple dynamical rules allow a detailed numerical study of current fluctuations.

In Appendix \ref{BDpredic} we apply the additivity formalisms of the previous section to study current fluctuations in 
the KMP model. In particular, we use eqs. (\ref{ldf1}) and (\ref{optprof}) to derive analytical expressions for the current LDF 
${\cal G}(q)$ and the associated optimal profiles $T_q(x)$, see Figs. \ref{predic}-\ref{predicprof}. In this case it can be shown that optimal profiles can be either monotone or 
non-monotone with a single maximum, see Appendix \ref{BDpredic} for the explicit calculations. In what follows, we compare this set of analytical predictions 
with computer simulation results.

\section{Numerical Test of the Additivity Principle}
\label{numres}

\begin{figure}
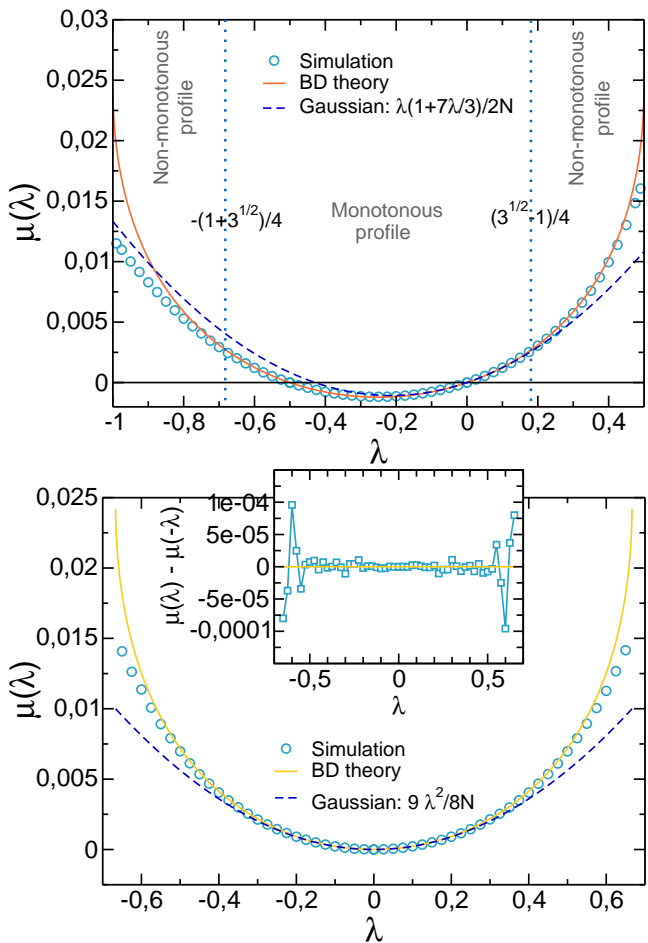

\centerline{\psfig{file=current-kmp-large-deviation.eps,width=8.5cm,clip}}
\centerline{\psfig{file=current-kmp-large-deviation-EQUILIBRIUM.eps,width=8.5cm,clip}}
\caption{
(Color online) Legendre transform of the current LDF for the KMP model in one dimension in a temperature gradient (top, $T_L=2$, $T_R=1$) and 
in equilibrium (bottom, $T_L=1.5=T_R$). Symbols correspond to numerical simulations, full lines to BD theory, and dashed lines to 
Gaussian approximations (see text). Errorbars (with 5 standard deviations) are always smaller than symbol sizes.
The vertical dotted lines in top panel signal the transition between deviations for which the associated
temperature profile is monotone (inner region) or non-monotone (outer region). In equilibrium profiles are non-monotone for all current fluctuations.
The inset in the bottom panel tests the Gallavotti-Cohen relation in equilibrium by plotting the difference $\mu(\lambda)-\mu(-\lambda)$.}
\label{ldfa}
\end{figure}

\begin{figure}
\centerline{\psfig{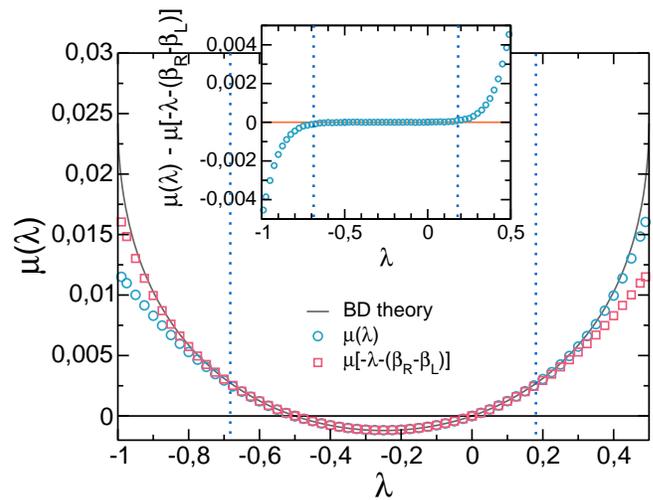}}
\caption{
(Color online) Measured $\mu(\lambda)$ and $\mu(-\lambda - {\cal E})$ superimposed. The Gallavotti-Cohen symmetry is
satisfied for a wide range of $\lambda$. The inset shows the difference $\mu(\lambda)-\mu(-\lambda - {\cal E})$.}
\label{ldfb}
\end{figure}

The simplicity and versatility of the KMP model allows us to obtain explicit analytical expressions for ${\cal G}(q)$ and $T_q(x)$ based on the additivity conjecture, 
see Appendix \ref{BDpredic}. Figs. \ref{predic} and \ref{predicprof} show the theoretical current LDF and the associated optimal profiles, respectively. 
We find that $\text{P}_N(q,T_L,T_R,t)$ is Gaussian around $\la q \ra$ with variance $\sigma(T)$, while 
non-Gaussian, exponential tails develop far from $\la q \ra$, with decay rates given by the inverse bath temperatures. 
Exploring by standard simulations these tails to check BD theory is very difficult,
since LDFs involve by definition exponentially-unlikely rare events. 
This is corroborated in Appendix \ref{stand}, where ${\cal G}(q)$ is measured directly but we are unable to gather enough statistics in 
the tails of the current distribution to validate or falsify the additivity hypothesis. 
Recently  Giardin\`a, Kurchan and Peliti \cite{sim} have introduced an efficient method to measure LDFs in many particle 
systems, based on a modification of the dynamics so that the rare events responsible of the large deviation are 
no longer rare \cite{sim2}. This method yields the Legendre transform of the current LDF, $\mu(\lambda)$, see eq. (\ref{mulamb}),
If $U_{C' C}$ is the transition rate from configuration $C$ to $C'$ of the associated stochastic process, the modified dynamics is defined as 
$\tilde{U}_{C' C}(\lambda)=U_{C' C} \exp(\lambda J_{C' C})$, where $J_{C' C}$ is the elementary current 
involved in the transition $C\to C'$.
It can be then shown (see Appendix \ref{algo}) that the natural logarithm of the largest eigenvalue of \emph{matrix} $\tilde{U}(\lambda)$ 
gives $\mu(\lambda)$. The method of Ref. \cite{sim} thus provides a way to measure $\mu(\lambda)$ 
by evolving many copies or clones of the system using the modified dynamics $\tilde{U}(\lambda)$, see Appendix \ref{algo}.

We applied the method of Giardin\`a \emph{et al.} to measure $\mu(\lambda)$ for the 1D KMP model with $N=50$, $T_L=2$ and $T_R=1$, see Fig. \ref{ldfa}, top panel. 
The agreement with BD theory is excellent for a wide $\lambda$-interval, say $-0.8<\lambda<0.45$, which corresponds to a very large range 
of current fluctuations, see inset to Fig. \ref{kpredic} in Appendix \ref{algo}. Moreover, the deviations observed for extreme current fluctuations 
are due to known limitations of the algorithm \cite{Pablo,sim,sim2,Pablo2}, so no violations of additivity are observed.
Notice that the spurious differences seem to occur earlier for currents \emph{against the gradient}, i.e. $\lambda <0$. 
In fact, we can use the Gallavotti-Cohen symmetry, which in $\lambda$-space now reads
$\mu(\lambda)=\mu(-\lambda-{\cal E})$ with ${\cal E}= (T_R^{-1}-T_L^{-1})$, to bound the range of validity of the algorithm: Violations of the fluctuation relation
indicate a systematic bias in the estimations provided by the method of Ref. \cite{sim}, see also \cite{Pablo2}. 
Fig. \ref{ldfb}. shows that the Gallavotti-Cohen symmetry holds in the large current interval for which the additivity principle predictions agree with measurements,
thus confirming its validity in this range.
However, we cannot discard the possibility of an additivity breakdown for extreme current fluctuations due to the onset of time-dependent 
optimal profiles expected in general in HFT \cite{Bertini}, although we stress that such scenario is not observed here. 

We also measured the current LDF in canonical equilibrium, i.e. for $T_L=T_R=1.5$, see the bottom panel in Fig. \ref{ldfa}. 
The agreement with BD theory is again excellent within the range of validity of our measurements, which expands a wide current interval, see inset to Fig. \ref{ldfa}, and
the fluctuation relation is verified except for extreme currents deviations, where the algorithm fails to provide reliable results.
Notice that, both in the presence of a temperature gradient and in canonical equilibrium, $\mu(\lambda)$ is parabolic around $\lambda=0$ meaning that current fluctuations
are Gaussian for $q\approx \la q\ra$, as demanded by the central limit theorem, see eqs. (\ref{Gsmallq})-(\ref{musmalllambda}) in Appendix \ref{BDpredic}. This observation 
is particularly interesting in equilibrium, where canonical and microcanonical ensembles behave differently (see below).

\begin{figure}
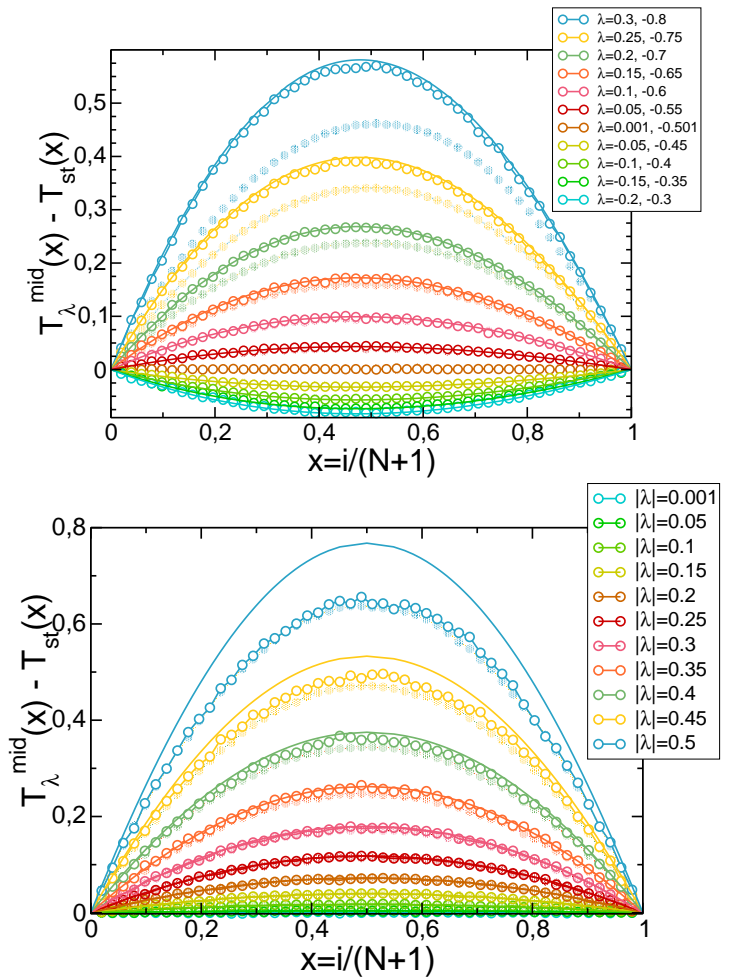

\centerline{\psfig{file=profiles-kmp-large-deviation-MIDTIME-OK.eps,width=9.cm,clip}}
\centerline{\psfig{file=profiles-kmp-large-deviation-EQUILIBRIUM-MID-OK.eps,width=9.5cm,clip}}
\caption{(Color online) Excess temperature profiles for different current fluctuations ($\bigcirc$), for a system subject to a temperature gradient (top, $T_L=2$, $T_R=1$) and 
in equilibrium (bottom, $T_L=1.5=T_R$). In all cases, agreement with BD theoretical predictions (lines) is very good within the range of validity of the computational method. 
Dotted symbols correspond to midtime profiles obtained from endtime statistics (see text).
}
\label{profmid}
\end{figure}

The additivity principle leads to the minimization of a functional of the temperature profile, $T_q(x)$, see  eqs. (\ref{ldf1}) and (\ref{optprof}).
A relevant question is whether this optimal profile is actually observable. We naturally define $T_q(x)$ as the average \emph{energy} 
profile adopted by the system during a large deviation event of (long) duration $t$ and time-integrated current $q t$, measured at an 
\emph{intermediate time} $1\ll \tau \ll t$, i.e. $T_q(x)\equiv T_q^{\text{mid}}(x)$ . 
Fig. \ref{profmid} shows the measured $T_{\lambda}^{\text{mid}}(x)$ for both the equilibrium and nonequilibrium settings,
and the agreement with BD predictions is again very good in all cases, with discrepancies appearing only 
for extreme  current fluctuations, as otherwise expected. See also Fig. \ref{profstand} in Appendix \ref{stand}.
This confirms the idea that the system indeed modifies its temperature profile to facilitate the deviation of the current, validating
the additivity principle as a powerful conjecture to compute both the current LDF and the associated optimal profiles.
Our numerical results show also that optimal profiles are indeed independent of the sign of the current, $T_{\lambda}(x)=T_{-\lambda-{\cal E}}(x)$ 
or equivalently $T_q(x)=T_{-q}(x)$, a counter-intuitive symmetry resulting from the reversibility of microscopic dynamics.
Notice that in the equilibrium case ($T_L=T_R$) optimal temperature profiles are always non-monotone with a single maximum for any current fluctuation $q\neq \la q \ra$
(the stationary profile is obviously flat). This is in stark contrast to the behavior predicted for current fluctuations in 
\emph{microcanonical} equilibrium, i.e. for a one-dimensional closed diffusive system on a ring \cite{Bertini,tprofile,Pablo3}. In this case the optimal profiles remain flat and 
current fluctuations are Gaussian up to a critical current value, at which profiles become time-dependent (traveling waves) \cite{Pablo3}. 
Hence current statistics can differ considerably depending on the particular equilibrium ensemble at hand, despite their equivalence for average 
quantities in the thermodynamic limit. Finally, notice also that equilibrium optimal profiles are symmetric with respect to $x=1/2$, as expected
since $T_L=T_R$.

\begin{figure}[t]
\centerline{\psfig{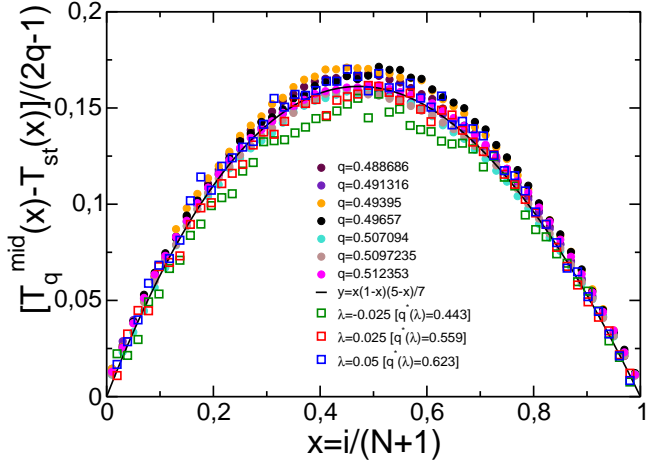}}
\caption{(Color online) Scaling plot of the excess profiles for small current fluctuations. Here we plot results obtained from standard
simulations (solid circles) and the advanced algorithm of Ref. \cite{sim} (open squares), as well as the theoretical prediction (line).}
\label{limit}
\end{figure}

For small enough current fluctuations around the average, $q\approx \la q\ra$ with $\la q \ra = 1/2$ for $T_L=2$ and $T_R=1$, BD theory predicts the limiting behavior
\begin{equation}
\frac{T_q(x)-T_{\text{st}}(x)}{2q-1} = \frac{1}{7} x (1-x) (5-x) + {\cal O}(2q-1) \, .
\label{limiting}
\end{equation}
Fig. \ref{limit} confirms this scaling for $T_q(x)$ and many different small current fluctuations around the average. 
In particular,  it shows data obtained both from standard simulations (see Appendix \ref{stand}) and using the advanced method of Ref. \cite{sim}. 

It is also interesting to study the statistics of configurations both during a large deviation event and at the end. They differ due to final transient effects 
which decay exponentially fast, but a connection exists between both regimes which highlights the symmetry of midtime statistics
resulting from the reversibility of microscopic dynamics (a symmetry akin to the fluctuation relation). Reversibility in stochastic dynamics 
stems from the condition of local detailed balance \cite{LS}, which implies a
relation between the forward modified dynamics for a current fluctuation, $\tilde{U}(\lambda)$, 
and the time-reversed modified dynamics for the negative fluctuation,
$\tilde{U}^{T}(-\lambda-{\cal E})$, see eq. (\ref{transpose}) in Appendix \ref{profil}. This  can be used to derive a
relation between midtime and endtime statistics (see Appendix \ref{profil}),
\be 
P_{\lambda}^{\text{mid}}(C) =A \, \frac{P_{\lambda}^{\text{end}}(C) P_{-\lambda-{\cal E}}^{\text{end}}(C)}{p_C^{\text{eq}}} \, ,
\label{probmid}
\ee
Here $P_{\lambda}^{\text{end}}(C)$ [resp. $P_{\lambda}^{\text{mid}}(C)$] is the probability of configuration $C$ at the end (resp. at 
intermediate times) of a large deviation event with current-conjugate parameter $\lambda$, and 
$p_{C}^{\text{eff}}= \exp [-\sum_{i=1}^N \beta_i e_i ]$ is an effective weight for configuration 
$C=\{e_i, i=1\ldots N\}$, with $\beta_i=T_L^{-1} + {\cal E} \frac{i-1}{N-1}$, while $A$ is a normalization constant.
Eq. (\ref{probmid}) implies that configurations with a significant contribution to the average profile at intermediate times are 
those with an important probabilistic weight at the end of both the large deviation event and its time-reversed process. 
An important consequence of eq. (\ref{probmid}) is hence that $P_{\lambda}^{\text{mid}}(C)=P_{-\lambda-{\cal E}}^{\text{mid}}(C)$, or equivalently 
$P_q^{\text{mid}}(C)=P_{-q}^{\text{mid}}(C)$, so midtime statistics does not depend on the sign of the current. This implies in particular that 
$T_q^{\text{mid}}(x)=T_{-q}^{\text{mid}}(x)$, but also that all higher-order profiles $\la e^n(x)\ra_q$ and spatial 
correlations $\la e^n(x_1)\ldots e^n(x_m)\ra_q $ are independent of the current sign $\forall n,m$. 
\begin{figure}
\centerline{\psfig{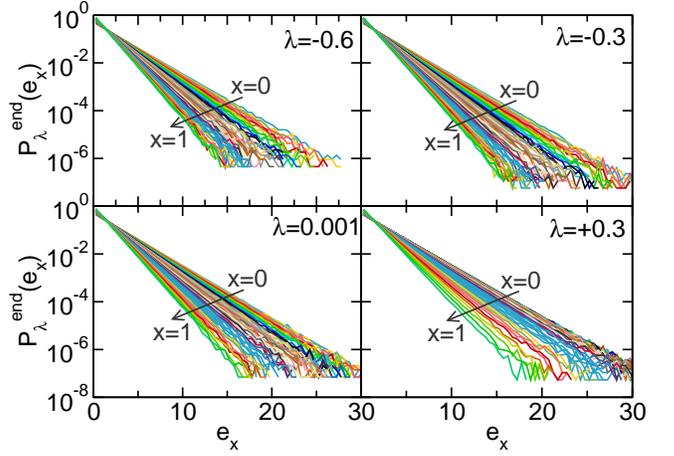}}
\caption{(Color online) Semilog plot of local energy histograms along the chain for different values of $\lambda$, at the end of the large deviation event.
Notice that, in all cases, energy distributions are very close to exponential.}
\label{histoloc}
\end{figure}

\begin{figure}
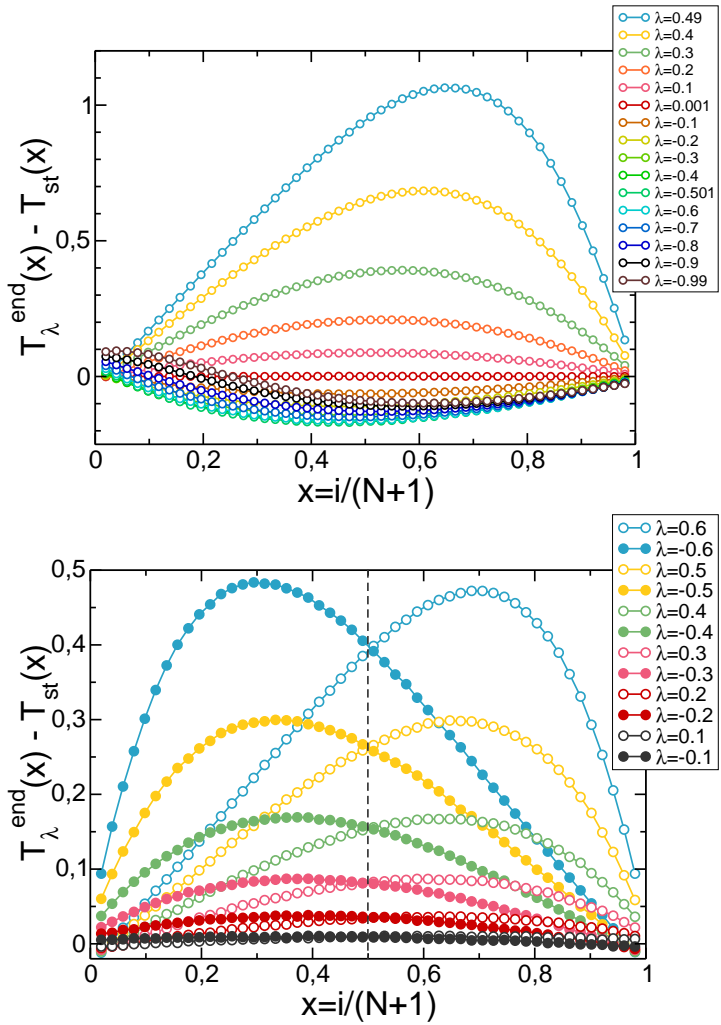

\centerline{\psfig{file=profiles-kmp-large-deviation.eps,width=9.5cm,clip}}
\centerline{\psfig{file=profiles-kmp-large-deviation-EQUILIBRIUM-END.eps,width=9.5cm,clip}}
\caption{(Color online) Excess temperature profiles measured at the end of the large deviation event for different values of $\lambda$, both in the presence of a temperature 
gradient (top), $T_L=2$ and $T_R=1$, and in canonical equilibrium (bottom), $T_L=1.5=T_R$. Notice that in all cases 
$T_{\lambda}^{\text{end}}(x) \neq T_{-\lambda-{\cal E}}^{\text{end}}(x)$, although for the equilibrium case the symmetry 
$T_{\lambda}^{\text{end}}(x)=T_{-\lambda}^{\text{end}}(1-x)$ is apparent.}
\label{profend}
\end{figure}

The above connection allows us to relate midtime and endtime profiles for a given current fluctuation. For that we need additionally a local equilibrium (LE) hypothesis, 
i.e. we now assume that spatial correlations at the end of a large deviation event are weak enough so the distribution $P_{\lambda}^{\text{end}}(C)$ can be approximately factorized, 
$P_{\lambda}^{\text{end}}(C)\approx \Pi_{i=1}^N P_{\lambda}^{\text{end}}(e_i)$. 
In this way we obtain a local equilibrium picture with local temperature parameter $T_{\lambda}^{\text{end}}(x=\frac{i}{N+1})$. 
This hypothesis can be numerically justified by measuring, at the end of the large deviation event, 
local energy distributions along the chain for different values of $\lambda$, see Fig. \ref{histoloc}. In all cases the distribution is compatible with local equilibrium 
to a large degree of accuracy. Using eq. (\ref{probmid}) and the LE hypothesis we thus find
\begin{equation}
T_{\lambda}^{\text{mid}}(x)= \frac{T_{\lambda}^{\text{end}}(x) \, T_{-\lambda-{\cal E}}^{\text{end}}(x)}{T_{\lambda}^{\text{end}}(x) + T_{-\lambda-{\cal E}}^{\text{end}}(x) - 
\beta_x \, T_{\lambda}^{\text{end}}(x) \, T_{-\lambda-{\cal E}}^{\text{end}}(x)} \, .
\label{profmid2}
\end{equation} 
Fig. \ref{profend} shows endtime profiles $T_{\lambda}^{\text{end}}(x)$ measured both in equilibrium (bottom) and nonequilibrium (top) conditions for different 
values of $\lambda$.  These profiles are clearly asymmetric upon current inversion, $T_{\lambda}^{\text{end}}(x) \neq T_{-\lambda-{\cal E}}^{\text{end}}(x)$, and 
most interestingly they show boundary resistance which depends on $\lambda$ and on the particular definition for the elementary current, 
see \cite{Pablo2}. In the equilibrium case the symmetry $T_q^{\text{end}}(x)=T_{-q}^{\text{end}}(1-x)$ resulting from the reflection invariance 
in this case ($T_L=T_R$) is apparent in Fig. \ref{profend} (bottom). 
Fig. \ref{profmid} also shows  midtime profiles obtained from the measured $T_{\lambda}^{\text{end}}(x)$ via eq. (\ref{profmid2}). 
The agreement with theoretical predictions and direct measurements of midtime profiles is good, though discrepancies appear for large enough 
current fluctuations, pointing out that corrections to LE are weak but increase for large current deviations. We show below that  these corrections 
are also present for small current fluctuations and can be measured.

\begin{figure}
\centerline{\psfig{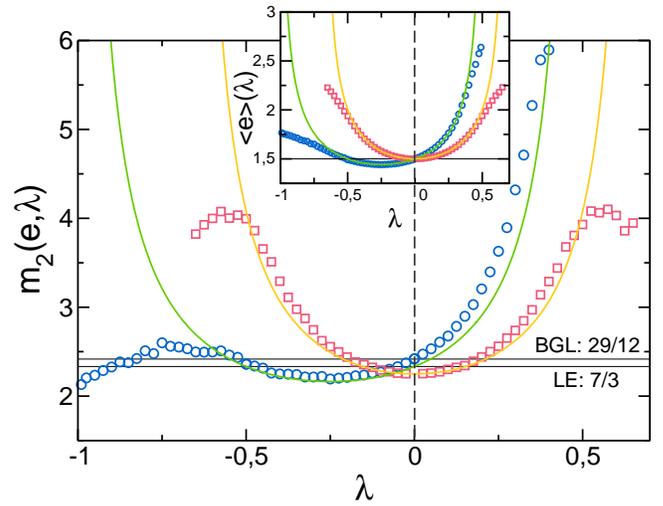}}
\caption{(Color online) Fluctuations of the total
energy per site versus $\lambda$ for both equilibrium ($\Box$, $T_L=1.5=T_R$) and nonequilibrium ($\bigcirc$, $T_L=2$, $T_R=1$) conditions. 
The lines stand for predictions based on the additivity principle plus a local equilibrium hypothesis. Inset: Average energy per site and BD prediction in both situations. 
Notice that, as before, deviations observed in all cases for extreme current fluctuations are spurious and result from known limitations of the method of Ref. \cite{sim}.}
\label{ener2}
\end{figure}

We can now explore physics beyond the additivity conjecture by studying fluctuations of the system total energy, $e(C)=N^{-1} \sum_{i=1}^N e_i$, 
for which current theoretical approaches cannot offer any general prediction. An exact result by Bertini, Gabrielli and Lebowitz (BGL) \cite{BGL} predicts that 
\be
m_2(e) = m_2^{LE}(e) + \frac{1}{12}(T_L-T_R)^2 \, ,
\label{m2}
\ee 
where $m_2(e)=N(\la e^2\ra -\la e\ra^2)$ is the variance of the total energy in the nonequilibrium steady state (NESS), 
$m_2^{LE}$ is the variance assuming a local equilibrium (LE) product measure,
and the last term 
reflects the correction to LE due to weak long-range correlations in the NESS \cite{BGL}, which in this case results in the enhancement of 
energy fluctuations. 
Corrections to LE vanish in the thermodynamic limit
but extend over macroscopic distances (of order $N$), giving rise in general to a non-local current LDF \cite{BGL}.
In our case, 
\be
m_2^{LE}=\frac{1}{3}\left(T_L^2 + T_LT_R + T_R^2\right)=\frac{7}{3} \approx 2.3333 \, , 
\label{m2LE1}
\ee
while $m_2=29/12 \approx 2.4166$. 
Fig. \ref{ener2} plots $m_2(e,\lambda)=N[\la e^2\ra_{\lambda} - \la e\ra_{\lambda}^2]$ as a function of $\lambda$ for both equilibrium and nonequilibrium conditions, 
showing a non-trivial, interesting structure which both BD theory and HFT cannot explain. One might obtain a theoretical prediction for $m_2(e,\lambda)$ 
by supplementing the additivity principle with a LE hypothesis, 
\be
P_{\lambda}(C)\propto \Pi_{i=1}^N \exp\left[-\frac{e_i}{T_{\lambda}(\frac{i}{N+1})}\right] \, , 
\label{LEapprox}
\ee
which results in
\be
m_2^{LE}(e,\lambda) = \int_0^1 \, T_{\lambda}(x)^2 \, dx \, . 
\label{m2LE}
\ee
This prediction agrees qualitatively with the observed behavior, though fine quantitative differences are apparent, see Fig. \ref{ener2}, as otherwise expected.
In particular we find that, out of equilibrium, $m_2^{LE}(e,0)\approx 2.33$ as corresponds to a LE picture, and 
in contrast to the measured value $m_2(e,0)=2.422 (14)$ in Fig. \ref{ener2}, which compares nicely with the exact BGL result $29/12$ (recall that $\lambda=0$ corresponds 
to $q=\la q\ra$). This shows that, even though LE is a sound numerical hypothesis to obtain $T_{\lambda}(x)$ from endtime statistics for small and moderate current fluctuations, 
see Fig. \ref{profmid} and eq. (\ref{profmid2}), corrections to LE become apparent at the fluctuating level even for small current fluctuations. 
This is also shown in Fig. \ref{enerstand} in Appendix \ref{stand}, where fluctuations of the total energy under nonequilibrium conditions are studied in standard simulations. 
On the other hand, in the canonical equilibrium case ($T_L=1.5=T_R$) no corrections to LE show up for $\lambda=0$ (i.e., for $q=\la q\ra=0$), as expected. 
However, as soon as $q\ne \la q\ra$,  deviations of $m_2(e,\lambda)$ from the LE prediction $m_2^{LE}(e,\lambda)$ are observed, thus showing that local equilibrium is
broken at the fluctuating level even for equal bath temperatures.

Finally, the inset to Fig. \ref{ener2} shows the average energy per site as a function of $\lambda$, together with the prediction
based on the additivity principle, $\la e\ra_{\lambda}= \int_0^1 T_{\lambda}(x) dx$. Agreement is again very good in the large range of currents explored. 
It is interesting to note that in order to sustain a current fluctuation above the average, $q>\la q\ra$ or equivalently $\lambda>0$, the nonequilibrium system ($T_L>T_R$) has always a
larger average energy than its equilibrium counterpart ($T_L=T_R$), while the reverse holds for current fluctuations below the average, $q<\la q\ra$, see inset to Fig. \ref{ener2}.

\section{Joint Fluctuations of the Current and the Profile}
\label{joint}

For long but finite times, the profile associated to a given current fluctuation is subject to fluctuations itself.
These joint fluctuations of the current and the profile are again not described by the additivity principle, but we may study them
by extending the additivity conjecture. In this way, we now assume that the probability to find a time-integrated current
$q/N$ \emph{and} a temperature profile $\bar{T}_q(x)$ after averaging for a long but finite time $t$ can be written as
\begin{equation}
W_N[\frac{q}{N},\bar{T}_q(x);t]\simeq \exp\left(+\frac{t}{N} \bar{\cal G}[q,\bar{T}_q(x)]\right)
\label{joint1}
\end{equation}
where now
\begin{equation}
\bar{\cal G}[q,\bar{T}_q(x)]=- \int_0^1 \frac{\left[q + \kappa[\bar{T}_q(x)]
    \bar{T}'_q(x) \right]^2}{2\sigma[\bar{T}_q(x)]} dx \, .
\label{joint2}
\end{equation}
Notice that here no minimization with respect to temperature profiles is performed, see eq. (\ref{ldf1}). 
In this scheme the profile obeying eq. (\ref{optprof}), i.e. the one which
minimizes the functional $\bar{\cal G}$, is the \emph{classical} profile
$T_{q}(x)$. For a given $q$ value we can make a perturbation of $\bar{T}_q(x)$
around its classical value, 
\begin{equation}
\bar{T}_q(x)=T_{q}(x)+\eta_q(x) \, .
\end{equation}
For large enough $t$, the joint probability of $q$ and $\eta_q(x)$ can be written as
\begin{equation}
\frac{W_N[q,\eta_q(x);t]}{\text{P}_N(q;t)}\simeq  \exp\left[-\frac{1}{2}\int dx dy
A_q(x,y)\eta_q(x)\eta_q(y) \right]
\end{equation}
where $\text{P}_N(q,t)$ is defined in equation (\ref{pn}), together with eqs.
(\ref{ldf1}) and (\ref{optprof}).
The integral kernel is
\begin{eqnarray}
\frac{N}{t}A_q(x,y) & = & \Big[
\frac{1}{2T_q^3}\frac{dT_q}{dx}\frac{d}{dx}-
\frac{1}{4T_q^2}\frac{d^2}{dx^2} \nonumber \\
& - & 2\frac{K(q)q^2}{T_q^2}\Big]\delta(x-y) \, .
\end{eqnarray}
\begin{figure}
\centerline{\psfig{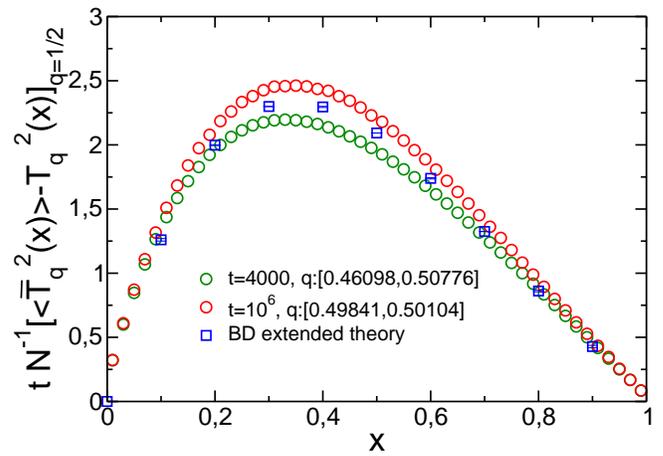}}
\caption{(Color online) Finite-time profile fluctuations $\la \eta_q^2(x)\ra$. Blue 
squares are the numerical evaluation of the series expansion from the extended BD theory (see text). Black
and red circles are standard simulation results for $q$'s in the interval
shown in the figure, $N=50$ and $t=4000$ and $t=10^6$.} 
\label{figure10}
\end{figure}
One can show that the kernel $A_q(x,y)$ is symmetric with respect to $x$ and $y$. In order to check the above
joint probability distribution, we studied the observable
\begin{equation}
\langle \bar{T}_q^2(x)\rangle-T_q^2(x)=\langle \eta_q^2(x)\rangle= A_q^{-1}(x,x) \, ,
\end{equation}
where
\begin{equation}
A_q^{-1}(x,y)=\sum_{n=1}^{\infty}\phi_n^{-1}v_{n}(x;q)v_n(y;q) \, ,
\end{equation}
and $v_n(x;q)$ and $\phi_n$ are the eigenvectors and eigenvalues of kernel
$A_q$, respectively,
\begin{equation}
\int dx A_q(x,y)v_n(x;q)=\phi_nv_n(y;q)\, , 
\end{equation}
with $v_n(0;q)=0=v_n(1;q)$. For $q=\la q \ra=1/2$ (nonequilibrium conditions, $T_L=2$, $T_R=1$) we were able to solve the
eigenvalue equation, yielding
\begin{eqnarray}
v_n(x;1/2)&=&B \, T_{1/2}(x)^{3/2}\biggl\{J_{-3/4}(\bar\phi_n T_L^2)J_{3/4}[\bar\phi_n T_{1/2}(x)^2]
\nonumber\\
&-&J_{3/4}(\bar\phi_nT_L^2)J_{-3/4}[\bar\phi_nT_{1/2}(x)^2]\biggr\} \, ,
\end{eqnarray}
where $\bar\phi_n=(\phi_nN/t)^{1/2}/(T_L-T_R)$, $J$'s are the Bessel functions and $B$
is the normalization factor that is obtained by requiring
\begin{equation}
\int_0^1 dx \, v_n(x;1/2)^2=1 \, .
\end{equation}
Finally, $\phi_n$ are the solutions of the equation
\begin{equation}
J_{3/4}(\bar\phi_n T_L^2)J_{-3/4}(\bar\phi_n T_R^2)=J_{-3/4}(\bar\phi_n
T_L^2)J_{3/4}(\bar\phi_n T_R^2) \, .
\end{equation}
We compare in Fig. \ref{figure10} the numerical evaluation of $A_{1/2}^{-1}(x,x)$ (where we have computed
$10$, $15$, $30$, $50$, $100$ and $200$ terms of the series and extrapolated to
$n\rightarrow\infty$) with the standard simulation results for $N=50$ and $t=4000$ and $t=10^6$. We
observe a good agreement between theoretical and simulation results. Notice
that  we average over a small $q$-window around $q=1/2$ in simulations. 
These results show that the BD functional $\bar{\cal G}[q,\bar{T}_q(x)]$ of eq. (\ref{joint2}) contains the
essential information on the joint fluctuations of the current and the 
average profile, extending the validity of the additivity principle to finite-time situations.

\section{Conclusions}
\label{conclu}

In this paper we have confirmed via extensive computer simulations the
validity of the additivity principle for current fluctuations in the 1D Kipnis-Marchioro-Pressuti
model of energy transport. In particular,
we found that the current distribution shows a
Gaussian regime for small current fluctuations and non-Gaussian, exponential
tails for large deviations of the current, such that in all cases the
fluctuation relation holds. We verified the existence of a well-defined
temperature profile associated to a given current fluctuation, different from
the steady-state profile and invariant under current reversal. In addition, we
extended the additivity conjecture to joint current-profile fluctuations.

Our results thus strongly support the additivity
hypothesis as an important tool to understand current statistics
in diffusive systems, opening the door to a general approach to a large class
of nonequilibrium phenomena
based on few simple principles. Our confirmation does not discard however the possible 
breakdown of additivity for extreme current fluctuations due to the onset of
time-dependent profiles, although we stress that this scenario is not 
observed here and would affect only the far tails of the current distribution. In this respect it would be interesting to study the KMP model on a 
ring, for which a dynamic phase transition to time-dependent profiles is expected \cite{Bertini,tprofile,Pablo3}. Also interesting is the possible extension of 
the additivity principle to low-dimensional systems with anomalous, non-diffusive transport properties \cite{we}, or to systems with several conserved 
fields or in higher dimensions.

\appendix

\section{Predictions using the Additivity Principle}
\label{BDpredic}

In this appendix we use the KMP model values for $\kappa(T)=\frac{1}{2}$ and $\sigma(T)=T^2$ in eqs. (\ref{ldf1}) and (\ref{optprof}) to derive explicit 
predictions for the current large deviation function in this model and the associated optimal temperature profiles. In what follows we assume 
$T_L>T_R$ without loss of generality. 
The differential equation for the optimal profile in the KMP case reads
\begin{equation}
\left(\frac{d T_q(x)}{dx}\right)^2 = 4q^2\left\{1+2T^2_q(x) K(q^2) \right\}  \, .
\label{optprofKMP}
\end{equation}
Here two different scenarios appear. On one hand, for large enough $K(q^2)$ the rhs of eq. (\ref{optprofKMP}) does not vanish $\forall x\in[0,1]$ and the resulting
profile is monotone. In this case, the optimal profile obeys
\be
\frac{d T_q(x)}{dx} = - 2|q| \sqrt{1+2T^2_q(x) K(q^2)}  \, .
\label{optprofKMPmonot}
\ee
On the other hand, for $K(q^2)<0$ the  rhs of eq. (\ref{optprofKMP}) may vanish at some points, resulting in a $T_q(x)$ that is non-monotone and takes 
an unique value $T_q^*\equiv \sqrt{-1/2K(q^2)}$ in the extrema. Notice that the rhs of the above equation may be written in this case as $4q^2[1-(T_q(x)/T_q^*)^2]$. It is then clear 
that, if non-monotone, the profile $T_q(x)$ can only have a single maximum $T_q(x^*)=T_q^*$ because: (i) $T_q(x)\le T_q^*$ $\forall x\in[0,1]$ for the 
profile to be a real function, and (ii) several maxima are not possible because they should be separated by a minimum, which is not allowed because of (i).
In this case
\begin{eqnarray}
\frac{d T_q(x)}{dx} \!=\! 
\left\{ \! \begin{array}{cc}
+2|q| \sqrt{1- \left(\frac{\displaystyle T_q(x)}{\displaystyle T_q^*}\right)^2} \, ,& 
\, {\displaystyle x<x^* } \\ \\
- 2|q| \sqrt{1-\left(\frac{\displaystyle T_q(x)}{\displaystyle T_q^*}\right)^2} \, ,& 
\, {\displaystyle x>x^* } 
\end{array}
\right.
\label{optprofKMPnomonot}
\end{eqnarray}
This leaves us with two separated regimes for current fluctuations, with the crossover happening for 
$|q|=\frac{T_L}{2}\left[\frac{\pi}{2} - \sin^{-1}\left(\frac{T_R}{T_L}\right) \right]$. This crossover current may be obtained from eq. (\ref{maxT}) below
by letting $T_q^*\to T_L$.

\subsection{Region I: $|q| < \frac{T_L}{2}\left[\frac{\pi}{2} - \sin^{-1}\left(\frac{T_R}{T_L}\right) \right]$}

In this region the optimal profile $T_q(x)$ is monotone in $x\in[0,1]$. Eq. (\ref{ldfmon}) then leads to
\begin{eqnarray}
{\cal G}(q) & = &\frac{q}{2} \left( \frac{1}{T_R} - \frac{1}{T_L}\right) - q^2 K(q^2) \\
 & + & \frac{|q|}{2}\left(\frac{\sqrt{1+2K(q^2)T_L^2}}{T_L} - 
\frac{\sqrt{1+2K(q^2)T_R^2}}{T_R}  \right) \, , \nonumber
\label{Gmonot}
\end{eqnarray}
where 
$K(q^2)$ is a constant defined by the boundary conditions.
The optimal temperature profile $T_q(x)$ in this regime is the solution of the following implicit equation
\be
2x|q|=\frac{1}{\sqrt{2K(q^2)}} \ln \left[ \frac{ T_L + \sqrt{T_L^2 + \frac{1}{2K(q^2)}}}{ T_q(x) + \sqrt{T_q(x)^2 + \frac{1}{2K(q^2)}}} \right]
\label{profmonot1}
\ee
whenever $K(q^2)>0$, or rather
\be
2x|q|=\frac{\sin^{-1}\left[\sqrt{-2K(q^2)}T_L\right] - \sin^{-1}\left[\sqrt{-2K(q^2)}T_q(x)\right]}{\sqrt{-2K(q^2)}} 
\label{profmonot2}
\ee 
in the case $-\frac{1}{2T_L^2}<K(q^2)<0$, see eq. (\ref{optprofKMPmonot}).
Making $x=1$ and $T_q(x=1)=T_R$ here, we obtain the implicit equation for the constant $K(q^2)$.

Some times it is interesting to work with the Legendre transform of the large deviation function, 
$\mu(\lambda)=N^{-1} \max_q \left[{\cal G}(q) + \lambda q\right] = 
{\cal G}(q_o) + \lambda q_o$, with $q_o(\lambda)$ given by $\partial_q {\cal G}(q_o) + \lambda=0$, and where now $-T_R^{-1} < \lambda < T_L^{-1}$. It then follows
\begin{equation}
\mu(\lambda)=-\frac{K(\lambda)}{N} [q_o(\lambda)]^2
\label{mumonot}
\end{equation}
where 
\be
2 |q_o(\lambda)|=\frac{1}{\sqrt{2K(\lambda)}} \ln \left[ \frac{ T_L + \sqrt{T_L^2 + \frac{1}{2K(\lambda)}}}{ T_R + \sqrt{T_R^2 + \frac{1}{2K(\lambda)}}} \right]
\label{qmonot1}
\ee
when $K(\lambda)>0$, or instead
\be
2 |q_o(\lambda)|=\frac{\sin^{-1}\left[\sqrt{-2K(\lambda)}T_L\right] - \sin^{-1}\left[\sqrt{-2K(\lambda)}T_R\right]}{\sqrt{-2K(\lambda)}}
\label{qmonot2}
\ee
in the case $-\frac{1}{2T_L^2}<K(\lambda)<0$,
and the constant $K(\lambda)\equiv K[q_o(\lambda)^2]$ is solution of the implicit equation
\begin{eqnarray}
\lambda & = & -\frac{1}{2}\left(\frac{1}{T_R} - \frac{1}{T_L} \right) \label{klambdamonot} \\
 & + & \frac{\text{sgn}\left[ q_o(\lambda)\right]}{2} \left[\frac{\sqrt{1+2 K(\lambda) T_R^2}}{T_R} - 
\frac{\sqrt{1+2 K(\lambda) T_L^2}}{T_L} \right] \nonumber
\end{eqnarray}
The optimal profile for a given $\lambda$ is just $T_{\lambda}(x) = T_{q_o(\lambda)}(x)$. In $\lambda$-space, monotone profiles are
expected for $\lambda\in [\lambda_- \, , \, \lambda_+]$ where $\lambda_{\pm} = -(T_R^{-1}-T_L^{-1})/2 \pm \sqrt{1-(T_R/T_L)^2}/(2T_R)$.

\subsection{Region II: $|q| > \frac{T_L}{2}\left[\frac{\pi}{2} - \sin^{-1}\left(\frac{T_R}{T_L}\right) \right]$}

In this case the optimal profile is non-monotone with a single maximum $T_q^*=T_q(x^*)$, see eq. (\ref{optprofKMPnomonot}). 
In this regime $K(q^2)<0$, and $T_q^*=1/\sqrt{-2K(q^2)}$. It follows
\begin{eqnarray}
&&{\cal G} (q) = \frac{|q|}{4T_q^*} \left[\pi - \sin^{-1}\left(\frac{T_R}{T_q^*} \right) - \sin^{-1}\left(\frac{T_L}{T_q^*} \right)\right]  \\
&+& \frac{q}{2} \left( \frac{1}{T_R} - \frac{1}{T_L}\right) - \frac{|q|}{2}\frac{\sqrt{1-\left( \frac{T_R}{T_q^*} \right)^2}}{T_R} 
- \frac{|q|}{2}\frac{\sqrt{1-\left( \frac{T_L}{T_q^*} \right)^2}}{T_L}  \nonumber \, .
\label{Gnomonot}
\end{eqnarray}
The optimal profile solution of eq. (\ref{optprofKMPnomonot}) is given by 
\begin{eqnarray}
x \!=\! 
\left\{ \! \begin{array}{cc}
{\displaystyle \frac{T_q^*}{2|q|} \left[ \sin^{-1} \left( \frac{T(x)}{T_q^*} \right) - \sin^{-1} \left( \frac{T_L}{T_q^*} \right) \right] } \, ,& 
\, {\displaystyle x<x^* } \\ \\
{\displaystyle 1 + \frac{T_q^*}{2|q|} \left[ \sin^{-1} \left( \frac{T_R}{T_q^*} \right) - \sin^{-1} \left( \frac{T(x)}{T_q^*} \right) \right]  } \, ,& 
\, {\displaystyle x>x^* } 
\end{array}
\right.
\label{profnomonot}
\end{eqnarray}
At the location of the profile maximum, $x=x^*$, both branches in the above equation must coincide and this condition provides equations for both $x^*$ and $T_q^*$
\begin{eqnarray}
|q| & = & \frac{T_q^*}{2} \left[ \pi - \sin^{-1} \left(\frac{T_L}{T_q^*}\right) - \sin^{-1} \left(\frac{T_R}{T_q^*}\right) \right] \label{maxT} \\
x^* & = & \frac{\displaystyle \frac{\pi}{2} - \sin^{-1} \left(\frac{T_L}{T_q^*} \right) }{\displaystyle \pi - \sin^{-1} \left(\frac{T_L}{T_q^*} \right) - 
\sin^{-1} \left(\frac{T_R}{T_q^*} \right)} 
\label{maximum}
\end{eqnarray}

As in Regime I, we find for the Legendre transform $\mu(\lambda)=-N^{-1} K(\lambda) q_o(\lambda)^2 = (2N)^{-1} [q_o(\lambda)/T_{\lambda}^*]^2$, with
$q_o(\lambda)$ defined in eq. (\ref{maxT}), $T_{\lambda}^*\equiv T_{q_o(\lambda)}^*$, 
and $\lambda$ given as in eq. (\ref{klambdamonot}) but with the notation change $K(\lambda)\to -1/[2 (T_{\lambda}^*)^2]$.
Non-monotone profiles are then expected for $\lambda \in [-T_R^{-1} , \lambda_-) \cup (\lambda_+ , T_L^{-1}]$.
\begin{figure}
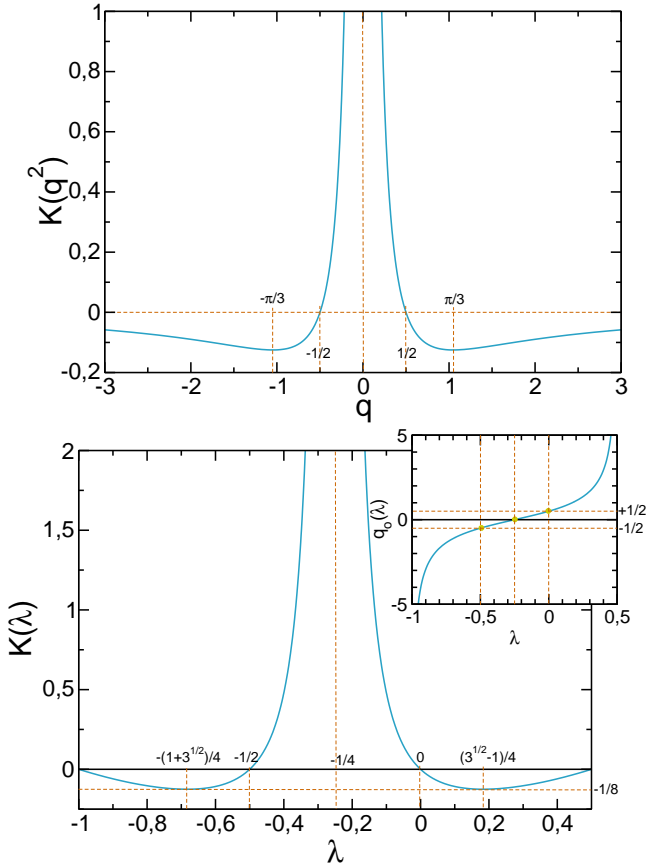

\centerline{\psfig{file=K-vs-q.eps,width=8cm,clip}}
\centerline{\psfig{file=K-vs-lambda.eps,width=8.5cm,clip}}
\caption{(Color online) Top panel: Constant $K$ as a function of q for $T_R=2$ and $T_L=1$. Bottom panel: the same constant as a function of $\lambda$.
The inset shows the current $q_o$ conjugated to $\lambda$.}
\label{kpredic}
\end{figure}

Fig. \ref{predic} in the main text shows the predicted ${\cal G}(q)$ for the KMP model. Notice that the large deviation function is zero for 
$q=\langle q \rangle = (T_L-T_R)/2$, and negative elsewhere. Moreover, for large current fluctuations it decays linearly, ${\cal G}(q) \to -q/T_{R,L}$ 
for $|q| \gg \langle q \rangle$.
For a small positive current fluctuation, $K(q^2) \to 0$ and 
\begin{equation}
{\cal G}(q) \approx - \frac{\displaystyle 3\left(|q|-\frac{T_L-T_R}{2} \right)^2}{2(T_L^2+T_LT_R+T_R^2)}  \, ,
\label{Gsmallq}
\end{equation}
which translates into 
\begin{equation}
\mu(\lambda)\approx \frac{\lambda}{2 N} \left[(T_L-T_R) + \frac{\lambda}{3}(T_L^2 + T_L T_R + T_R^2) \right] \, ,
\label{musmalllambda}
\end{equation}
for the Legendre transform. Therefore the probability of small current fluctuations is Gaussian in $q$ while it 
becomes exponential for large enough deviations from the average, see eq. (\ref{pn}). It is easy to show that the Gallavotti-Cohen symmetry holds, with
\begin{equation}
{\cal G}(q) - {\cal G}(-q) = 2 q \int_{T_R}^{T_L} \frac{\kappa(T)}{\sigma(T)} = q \left(\frac{1}{T_R} - \frac{1}{T_L} \right) \, ,
\label{GCpredic}
\end{equation}
or equivalently 
\begin{equation}
\mu(\lambda) = \mu( -\lambda - {\cal E}) \, ,
\label{GCmupredic}
\end{equation}
with ${\cal E}\equiv (T_R^{-1}-T_L^{-1})$.
Fig. \ref{predicprof} in the main text shows the optimal temperature profiles for different current deviations.
Notice that the optimal profile is independent of the sign of the current, i.e. $T_q(x)=T_{-q}(x)$, reflecting the 
time-reversal symmetry of microscopic dynamics \cite{GC,LS}.
Finally, Fig. \ref{kpredic} shows, for information purposes, the integration constant $K$ as a function of both $q$ and $\lambda$, as well as 
$\lambda$-dependence of $q_o(\lambda)$.

\section{Standard Simulations}
\label{stand}

\begin{figure}
\centerline{\psfig{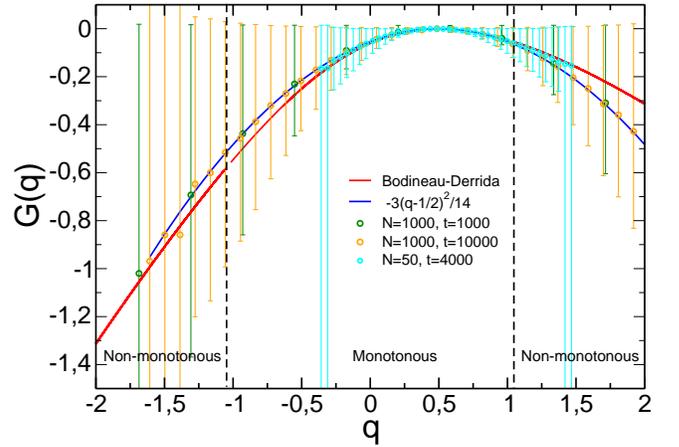}}
\caption{(Color online) ${\cal G}(q)$ measured for different system sizes $N$ and measurement times $t$ (see text), with $T_L=2$ 
and $T_R=1$ fixed. Lines correspond to BD theory and the Gaussian approximation. }
\label{standardldf}
\end{figure}

In order to see how far standard simulations can go in evaluating current large fluctuations, and to cross-check our results
with the more advanced simulation methods described in Appendix \ref{algo},
we performed a large number of steady-state simulations of long duration $t$, with $T_L=2$ and $T_R=1$, measuring the total time-integrated 
current $Q_{t}=qt$ and accumulating statistics for $q$. Fig. \ref{standardldf} shows the measured
${\cal G}(q)$ obtained for different system sizes $N$ and durations $t$. 
Our simulations for $N=1000$ and different times $t<N^2$ follow closely the Gaussian law ${\cal G}(q)\approx -3(q-1/2)^2/14$ 
obtained from the first two moments prescribed by the additivity principle in this case, namely
\begin{eqnarray}
m_1 & = &\frac{T_L-T_R}{2} \nonumber \\
m_2 & = & \frac{T_L^2+T_LT_R+T_R^2}{3} \nonumber \, .
\label{moments}
\end{eqnarray}
This Gaussian behavior is expected for small fluctuations around the average current, see eq. (\ref{Gsmallq}), but deviations 
away from Gaussianity should be already observed in the current range studied, see the theoretical prediction. In particular, the theoretical ${\cal G}(q)$ 
implies a nonzero third central moment, but we have not found numerical evidence of such a deviation for $N=1000$. 
This lack of structure stems from the relatively short duration of the simulations for $N=1000$, i.e. 
our results are not in the diffusive regime ($t<N^2$ here) and therefore we have not reached the asymptotic behavior.

\begin{figure}
\centerline{\psfig{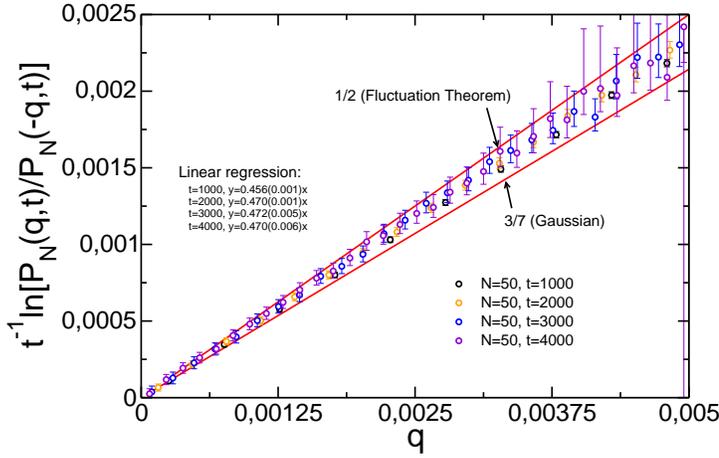}}
\caption{(Color online) Test of the fluctuation theorem of Gallavotti and Cohen. Here we explore $N=50$ and different maximum times $t$. If BD theory 
holds a slope $1/2$ is expected, while Gaussian behavior involves a slope $3/7$.}
\label{standardGC}
\end{figure}

We performed two set of simulations in the diffusive regime $t> N^2$, namely $N=50$ with $t=10^6$ and $t=4000$. In the first case there were no events
outside the current interval $q\in[0.45,0.56]$, for which the BD prediction is numerically indistinguishable from the Gaussian one. On 
the other hand, the case $N=50$ and $t=4000$ shows systematic deviations from Gaussian behavior, seemingly compatible with BD theory, see 
Fig. \ref{standardldf}. However, large errorbars resulting from the difficulty of gathering statistics in this rare-fluctuation regime do not allow us to exclude Gaussian 
behavior. In this way, standard simulation results are inconclusive, as otherwise expected, and the more refined simulation techniques of Appendix \ref{algo} are called 
for.

We also tested the Gallavotti-Cohen relation in standard simulations for our system. This symmetry implies that
\begin{equation}
\lim_{t \to \infty} \frac{1}{t} \ln \frac{P_N(q,T_L,T_R,t)}{P_N(-q,T_L,T_R,t)} = {\cal E} \, q \, ,
\label{GC}
\end{equation}
where ${\cal E}=(T_R^{-1}-T_L^{-1})=1/2$ in this case. Notice that if we assume $P_N(q,T_L,T_R;t)$ to be Gaussian with the moments 
defined above, then one expects ${\cal E}=3/7$. Fig. \ref{standardGC} shows the above quotient as measured for $N=50$ 
and different values of $t$. It shows a systematic deviation from Gaussian behavior which increases with $t$. However, we
do not see clearly ${\cal E}=1/2$, and this means again that our standard simulations are still far from the true asymptotic regime in $t$. 

\begin{figure}
\centerline{\psfig{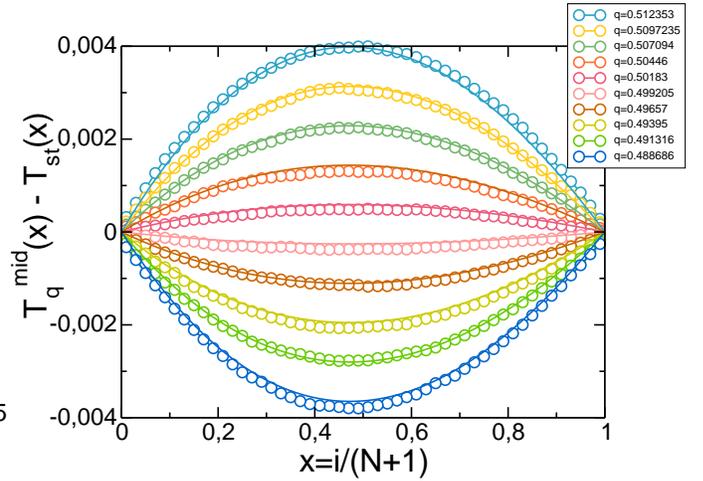}}
\caption{(Color online) Excess average profiles during a large deviation event for small current fluctuations, as measured 
in standard simulations. Agreement with BD theoretical predictions (lines) is excellent.}
\label{profstand}
\end{figure}

Another prediction of the additivity principle concerns the existence of an optimal temperature profile that the system adopts in order to facilitate
a given current fluctuation. We measured in standard simulations the average energy profile during a current large deviation event, obtaining the 
results shown in Fig. \ref{profstand}. As above, only for small current fluctuations we could gather enough statistics for the data to be significative. 
In any case, the theoretical optimal profiles compare nicely with data, confirming the existence of a well-defined temperature profile for each current deviation. 

\begin{figure}[t]
\centerline{
\psfig{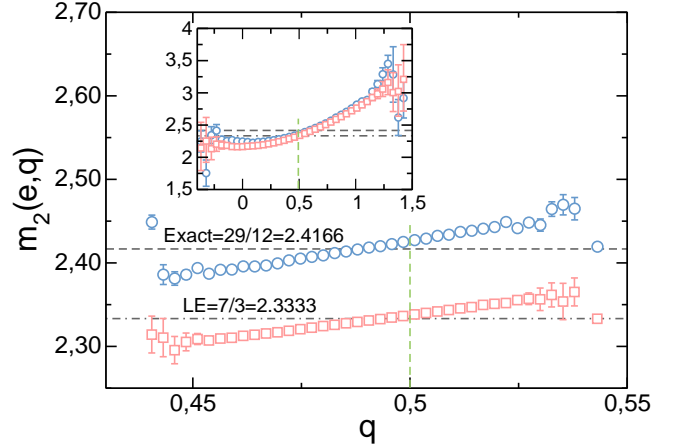}}
\caption{(Color online) Fluctuations of total energy vs $q$ measured in standard simulations for $t=10^6$ ($\bigcirc$) and LE results ($\Box$).
Inset: Similar results for $t=4000$. Notice the non-trivial structure.
}
\label{enerstand}
\end{figure}

We also measured the fluctuations of the total energy in standard simulations. Fig. \ref{enerstand} shows our results in this case. In particular, we measured
$m_2(e)=2.4 \, (1)$ for $N=50$ and a maximum time $t=4000$ and $m_2(e)=2.42 \, (2)$ for $t=10^6$, in agreement with eq. (\ref{m2LE1}). This figure also shows
$m_2(e,q)$ and $m_2^{LE}(e,q)$ build from simulation data for $T_q(x)$. As in Fig. \ref{ener2}, we see a clear deviation from local equilibrium and a
well defined structure not predicted by BD theory. Notice that, again, values of $m_2(e,q)$ for $q=1/2$ coincide with the expected average values with no current constraint.
The data shown in this figure agree nicely with those measured with the advanced technique in the studied range, see Fig. \ref{ener2}.

\section{Evaluation of Large-Deviation Functions}
\label{algo}

Large deviation functions are very hard to measure in experiments or simulations because they involve by definition
exponentially-unlikely events, see eq.(\ref{pn}). Recently, Giardin\`a, Kurchan and Peliti \cite{sim} have introduced an efficient algorithm to measure 
the probability of a large deviation for observables such as the current or density in stochastic many-particle systems. 
The algorithm is based on a  modification of the underlying stochastic dynamics so that the rare events responsible of the large deviation are no longer rare,
and it has been extended for systems with continuous-time stochastic dynamics \cite{sim2}. 
Let $U_{C' C}$ be the transition rate from configuration $C$ to $C'$. The probability 
of measuring a time-integrated current $Q_t$ after a time $t$ starting from a configuration $C_0$ can be written as
\begin{equation}
P(Q_t,t;C_0) = \sum_{C_{t}..C_1} U_{C_t C_{t-1}}..U_{C_1 C_0} 
\, \delta (Q_t - \sum_{k=0}^{t-1}J_{C_{k+1} C_k}) \, ,
\label{recurr1}
\end{equation}
where $J_{C' C}$ is the elementary current involved in the transition $C\to C'$. For long times we expect the information on the initial 
state $C_0$ to be lost, $P(Q_t,t;C_0) \to P(Q_t,t)$. In this limit $P(Q_t,t)$ obeys the usual large deviation principle 
$P(Q_t,t)\sim \exp[+t {\cal F}(q=Q_t/t)]$. In most cases it is convenient to work with the moment-generating function of the above distribution
\begin{eqnarray}
\Pi(\lambda,t) &=& \sum_{Q_t} \text{e}^{\lambda Q_t} P(Q_t,t)  \\
&=& \sum_{C_{t}..C_1} U_{C_t C_{t-1}}..U_{C_1 C_0} \, \text{e}^{\lambda  \sum_{k=0}^{t-1}J_{C_{k+1} C_k}} \, . \nonumber
\label{pi1}
\end{eqnarray}
For long $t$, we have {$\Pi(\lambda,t) \to \exp[+t \mu(\lambda)]$, with $\mu(\lambda)= \max_q [{\cal F}(q) + \lambda q]$}. We can
now define a modified dynamics, $\tilde{U}_{C' C}\equiv \text{e}^{\lambda J_{C' C}}\, U_{C' C}$, so
\begin{equation}
\Pi(\lambda,t)  = \sum_{C_{t}\ldots C_1} \tilde{U}_{C_t C_{t-1}} \ldots \tilde{U}_{C_1 C_0}  \, .
\label{pi2}
\end{equation}
This dynamics is however not normalized, $\sum_{C'} \tilde{U}_{C' C}\neq 1$.

We now introduce Dirac's bra and ket notation, useful in the context of the quantum Hamiltonian formalism for the master equation \cite{schutz,schutz2}, see also \cite{sim,Rakos}. 
The idea is to assign to each system configuration $C$ a vector $|C\ra$ in phase space, which together with its transposed vector $\la C |$, form 
an orthogonal basis of a complex space and its dual \cite{schutz,schutz2}. For instance, in the simpler case of systems with a finite number of available configurations
(which is not the case for the KMP model), one could write $|C\ra^T = \la C|=(\ldots 0 \ldots 0, 1, 0 \ldots 0 \ldots )$, i.e. all 
components equal to zero except for the component corresponding to configuration $C$, which is $1$. In this notation, 
$\tilde{U}_{C' C}= \la C' | \tilde{U} | C \ra$, and a probability distribution can be written as a probability vector
\begin{equation}
| P(t) \ra = \sum_C P(C,t) |C\ra \nonumber \, ,
\end{equation}
where $P(C,t)=\la C| P(t) \ra$ with the scalar product $\la C' | C\ra = \delta_{C'C}$. If $\la s|=(1\ldots 1)$, normalization then implies $\la s|P(t)\ra =1$.

\begin{figure}
\centerline{\psfig{file=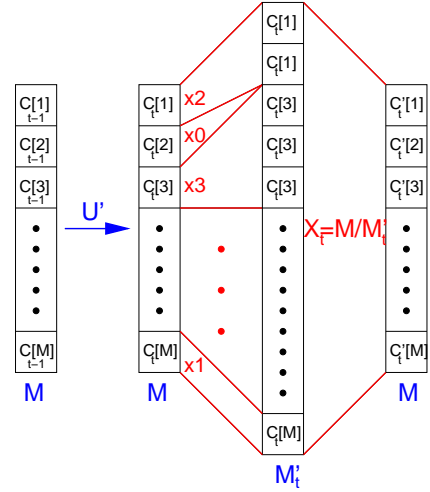,width=5.5cm,clip}}
\caption{(Color online) Sketch of the evolution and cloning of the copies during the evaluation of the large deviation function.}
\label{sketch}
\end{figure}

With the above notation, we can write the spectral decomposition
$\tilde{U}(\lambda)=\sum_j \text{e}^{\Lambda_j(\lambda)} |\Lambda_j^R(\lambda) \ra \la \Lambda_j^L(\lambda) |$, where we assume that a complete biorthogonal 
basis of right and left eigenvectors for matrix $\tilde{U}$ exists, $\tilde{U} |\Lambda_j^R(\lambda) \ra = \text{e}^{\Lambda_j(\lambda)} |\Lambda_j^R(\lambda) \ra$ and 
$\la \Lambda_j^L(\lambda)| \tilde{U} = \text{e}^{\Lambda_j(\lambda)} \la \Lambda_j^L(\lambda)|$. Denoting as $\text{e}^{\Lambda(\lambda)}$ the largest eigenvalue 
of $\tilde{U}(\lambda)$, with associated right and left eigenvectors $|\Lambda^R(\lambda) \ra$ and $\la \Lambda^L(\lambda)|$, respectively, and writing 
$\Pi(\lambda,t) = \sum_{C_t} \la C_t | \tilde{U}^t | C_0 \ra$, we find for long times
\begin{equation}
\Pi(\lambda,t) \xrightarrow{t \gg 1} \text{e}^{+t \Lambda(\lambda)}  \langle \Lambda^L(\lambda) | C_0 \rangle 
\left( \sum_{C_t} \langle C_t | \Lambda^R(\lambda) \rangle \right)\, .
\label{muasymp}
\end{equation}
In this way we have $\mu(\lambda)=\Lambda(\lambda)$, so the Legendre transform of the current LDF is
given by the natural logarithm of the largest eigenvalue of $\tilde{U}(\lambda)$. 
In order to evaluate this eigenvalue, and given that dynamics $\tilde{U}$ 
is not normalized, we introduce the exit rates $Y_C=\sum_{C'} \tilde{U}_{C' C}$, and define the normalized dynamics 
$U'_{C' C}\equiv Y_C^{-1} \tilde{U}_{C' C}$. Now
\begin{equation}
{\Pi (\lambda,t)=  \sum_{C_{t}\ldots C_1} Y_{C_{t-1}} U'_{C_t C_{t-1}} \ldots Y_{C_0} U'_{C_1 C_0} }
\label{pilambda}
\end{equation}
This sum over paths can be realized by considering an ensemble of $M \gg 1$ copies (or clones) of the system, evolving sequentially according 
to the following Monte Carlo scheme \cite{sim}:
\begin{enumerate}
\item[I] Each copy evolves independently according to modified normalized dynamics $U'_{C' C}$.
\item[II] Each copy $m\in [1,M]$ (in configuration $C_t[m]$ at time $t$) is cloned with rate $Y_{C_t[m]}$. This means that, for each copy $m\in [1,M]$, 
we generate a number $K_{C_t[m]}=\lfloor Y_{C_t[m]} \rfloor +1$ of identical clones with probability $Y_{C_t[m]} - \lfloor Y_{C_t[m]} \rfloor$, or
$K_{C_t[m]}=\lfloor Y_{C_t[m]} \rfloor$ otherwise (here $\lfloor x \rfloor$ represents the integer part of $x$). Note that if $K_{C_t[m]}=0$ the copy
may be killed and leave no offspring. This procedure gives rise to a total of $M'_t=\sum_{m=1}^M K_{C_t[m]}$ copies after cloning all of the original 
$M$ copies.
\item[III] Once all copies evolve and clone, the total number of copies $M'_t$ is sent back to $M$ by an uniform cloning probability $X_t=M/M'_t$.
\end{enumerate}
Fig. \ref{sketch} sketches this procedure. It then can be shown that, for long times, we recover $\mu(\lambda)$ via
\begin{equation}
\mu(\lambda)  = -\frac{1}{t} \ln \left(X_t \cdots X_0 \right)   \qquad \text{for } t\gg 1
\label{musim}
\end{equation}
To derive this expression, first consider the cloning dynamics above, but without keeping the total number of clones constant, i.e. forgetting about 
step III. In this case, for a given history $\{C_t,C_{t-1}\ldots C_1,C_0 \}$, the number ${\cal N}(C_t\ldots C_0,t)$ of copies in configuration 
$C_t$ at time $t$ obeys ${\cal N}(C_t\ldots C_0,t)=Y_{C_{t-1}} U'_{C_t C_{t-1}}  {\cal N}(C_{t-1}\ldots C_0,t-1)$, so that 
\begin{equation}
{\cal N}(C_t\ldots C_0,t)=Y_{C_{t-1}} U'_{C_t C_{t-1}} \ldots Y_{C_0} U'_{C_1 C_0} {\cal N}(C_0,0) \, . 
\label{fraccop}
\end{equation}
Summing over all histories of duration $t$, see eq. (\ref{pilambda}), we find that the average of the total number of clones at long times shows 
exponential behavior, $\la {\cal N} (t)\ra = \sum_{C_t\ldots C_1} {\cal N}(C_t\ldots C_0,t) \sim {\cal N}(C_0,0) \exp[+t \mu(\lambda)]$. Now, going back to 
step III above, when the fixed number of copies $M$ is large enough, we have $X_t = \la {\cal N} (t-1)\ra/\la {\cal N} (t)\ra$ for the global 
cloning factors, so $X_t \cdots X_1 = {\cal N} (C_0,0)/\la {\cal N} (t)\ra$ and we recover expression (\ref{musim}) for $\mu(\lambda)$.

In this paper we used the above method to measure the current LDF for the Kipnis-Marchioro-Presutti model in one dimension, described in Section \ref{model}. 
For this model the transition rate from a configuration $C=\{e_1\ldots e_N \}$ to another configuration  $C'_y=\{e_1\ldots e'_y,e'_{y+1}\ldots e_N \}$, with $y\in[0,N]$ and 
the pair $(e'_y,e'_{y+1})$ defined as in eqs. (\ref{defKMP1})-(\ref{defKMP2}), can be written as
\begin{eqnarray}
\label{rate}
U_{C'_y C} \!=\!
\left\{ \! \begin{array}{cc}
{\displaystyle (N+1)^{-1}} \, ,& \, {\displaystyle y\in[1,N-1] } \\ \\
{\displaystyle \frac{\beta_- \text{e}^{\beta_- e_1}}{N+1} \text{E}_1\left[\beta_- \max(e_1,e'_1)\right] } \, ,& \, {\displaystyle y=0 } \\ \\
{\displaystyle \frac{\beta_+ \text{e}^{\beta_+ e_N}}{N+1} \text{E}_1\left[\beta_+ \max(e_N,e'_N)\right] } \, ,& \, {\displaystyle y=N \, .}
\end{array}
\right. \nonumber
\end{eqnarray}
Here $\text{E}_1(x)=-\text{Ei}(-x)$, where $\text{Ei}(x)$ is the exponential integral function, or
\begin{equation}
\text{E}_1(x)=\int_x^{\infty} \text{d}u \frac{\text{e}^{-u}}{u} \, .
\label{expint}
\end{equation}
It appears when integrating over all possible pairs $(p,\tilde{e}_{L,R})$ that can result on a given $e'_{1,N}$, respectively, see eq. (\ref{defKMP2}) in Section \ref{model}. 
It is easy to show that $U_{C'_y C}$ is normalized as it should, so $\sum_{C'_y} U_{C'_y C} = 1$.

\begin{figure}
\centerline{\psfig{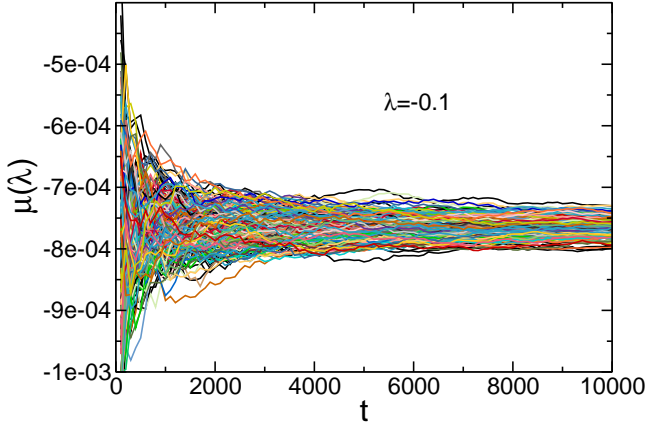}}
\caption{(Color online) Time evolution of $\mu(\lambda)$ for $\lambda=-0.1$ and many different initial conditions. Here $N=50$, $M=10^3$, and $T_L=2$, $T_R=1$.}
\label{ldftime}
\end{figure}

In order to measure current fluctuations we need to provide a microscopic definition of the energy current 
involved in an elementary move. There are many different ways to define this current: the energy exchanged per unit time with one of the boundary heat baths, 
the current flowing between two given nearest neighbors, or its spatial average, etc. Assuming that energy cannot accumulate 
in the system \emph{ad infinitum}\cite{Derrida,Derrida2,Rakos}, all these definitions give equivalent results for the current large deviation function 
in the long time limit. However, this is not so for some observables different from the large deviation function (e.g. for average profiles 
measured \emph{at the end} of the large deviation event; see Ref. \cite{Pablo2}). In our case, the following choice turns out to be convenient
\begin{eqnarray}
\label{current}
J_{C'_y C}  \!=\!
\left\{ \! \begin{array}{cc}
{\displaystyle \frac{e_y-e'_y}{N-1}} & \quad {\displaystyle y\in[1,N-1]  \text{ (bulk exchange) } } \\ \\
{\displaystyle 0 } &{\displaystyle y=0,N \quad \text{ (boundary baths) } }
\end{array}
\right.
\end{eqnarray}
That is, we measure the energy current flowing through the bulk of the system. Using this current definition and eq. (\ref{rate}), we may write the modified 
normalized dynamics $U'_{C'_y C}\equiv Y_C^{-1} U_{C'_y C} \, \exp[\lambda J_{C'_y C}]$, which for $y\in[1,N-1]$ reads
\be
U'_{C'_y C} = \frac{\displaystyle \text{e}^{\bar{\lambda} (e_y-e'_y)}}{Y_C(N+1)} \, ,
\label{moderate}
\ee
with $\bar{\lambda}=\lambda/(N-1)$, while $U'_{C'_y C}\equiv Y_C^{-1} U_{C'_y C}$ for $y=0,N$, see eq. (\ref{current}). The exit rate is given by
\begin{equation}
Y_C=\frac{2}{N+1} + \sum_{y=1}^{N-1} \frac{\displaystyle
  \text{e}^{\bar{\lambda} e_y} - \text{e}^{-\bar{\lambda}
    e_{y+1}}}{\displaystyle \bar{\lambda} (N+1) (e_y + e_{y+1})} \, .
\label{exitrate}
\end{equation}
In these paper we simulate a system of size $N=50$, with $T_L=2$ and $T_R=1$, using $M=10^3$ copies of the system and a maximum time of $t=10^4$ Monte Carlo steps. 
For a given initial condition, we average the measured $\mu(\lambda)$ for
different times once in the steady state, after a relaxation time of $2\times
10^3$ Monte Carlo steps. In addition, we 
average results over many independent initial conditions, in which local initial energies $e_i$ are randomly drawn according to the Gibbs distribution with temperature 
parameter $T_{\text{st}}[x=i/(N+1)]$ corresponding to the linear, steady
temperature profile. Fig \ref{ldftime} shows the convergence of $\mu(\lambda)$
in time for 
a given value of $\lambda$ and many different initial conditions. Using the
above method, we obtained an accurate measurement of the current large deviation 
function, see Fig. \ref{ldfa} in Section \ref{numres}.

\section{Time Reversibility and Statistics during a Large Fluctuation}
\label{profil}

In this Appendix we use the time reversibility of the underlying stochastic dynamics to study the system statistics during a large deviation event and the symmetries 
of the large deviation function and the associated optimal profiles, using the  formalism described in Appendix \ref{algo}. In particular, we
describe a relation between system statistics at the end of the large deviation event and for intermediate times. 
First, consider the probability $P(C_t,Q_t,t)$ that the system is in configuration $C_t$ at time $t$ with a 
total time-integrated current $Q_t$. As in the previous appendix, we drop the dependence of this probability on the initial state $C_0$, which we assume 
lost for long enough times. This probability obeys the following master equation
\begin{equation}
P(C_t,Q_t,t) = \sum_{C'} U_{C_t C'} P(C',Q_t-J_{C_t C'},t-1) \, .
\label{mastereq}
\end{equation}
which by iterating in time leads to
\begin{equation}
P(C_t,Q_t,t) = \sum_{C_{t-1}..C_1} U_{C_t C_{t-1}}..U_{C_1 C_0}\, \delta (Q_t - \sum_{k=0}^{t-1}J_{C_{k+1} C_k}) \, ,
\label{recurr}
\end{equation}
and it is clear that $P(Q_t,t)=\sum_{C_t} P(C_t,Q_t,t)$, see eq. (\ref{recurr1}) in the previous appendix. 
Now, $P_q^{\text{end}}(C_t)\equiv P(C_t,Q_t,t)/P(Q_t,t;C_0)$ is 
the probability of having a configuration $C_t$ \emph{at the end} of a large deviation event associated to a current $q=Q_t/t$. Defining
$\Pi(C_t,\lambda,t)=\sum_{Q_t} \exp(\lambda Q_t) P(C_t,Q_t,t)$ so that
\begin{equation}
\Pi(C_t,\lambda,t)  = \sum_{C_{t-1}\ldots C_1} \tilde{U}_{C_t C_{t-1}} \ldots \tilde{U}_{C_1 C_0} \, ,  
\label{PiCt}
\end{equation}
with $\tilde{U}_{C' C}(\lambda)= U_{C' C} \exp(\lambda J_{C' C})$, 
one can easily show that, for long times $t$, $P_{\lambda}^{\text{end}}(C_t)\equiv \Pi(C_t,\lambda,t)/\Pi(\lambda,t) = P_{q_o(\lambda)}^{\text{end}}(C_t)$, 
where $q_o(\lambda)$ is the current conjugated to parameter $\lambda$, and $\Pi(\lambda,t)$ is defined in eq. (\ref{pi2}). 
Using the spectral decomposition of  Appendix \ref{algo}, it is simple to show that $P_q^{\text{end}}(C_t)\propto \la C_t | \Lambda^R(\lambda)\ra$, 
so the right eigenvector $|\Lambda^R(\lambda)\ra$ associated to the largest eigenvalue of \emph{matrix} $\tilde{U}(\lambda)$ 
gives the probability of having any configuration \emph{at the end} of the large deviation event. Noticing that, for the Monte Carlo algorithm described 
in the previous appendix, the fraction of clones or copies in state $C_t$ is proportional to $\la C_t | \Lambda^R(\lambda)\ra$ for long times, see eq. (\ref{fraccop}), 
we deduce that the the average profile among the set of clones yields the mean temperature profile \emph{at the end} of the large deviation event, 
$T_{\lambda}^{\text{end}}(x)$.

The initial and final time regimes during a large deviation event show transient behavior which differs from the behavior in the \emph{bulk} of the large
deviation event, i.e. for intermediate times \cite{Derrida}. In particular, as we will show here, midtime and endtime statistics are different, though intimately related
as a result of the time reversibility of the microscopic dynamics.
Let $\bar{P}(C_{\tau},\lambda,\tau,t)$ be the probability that the system was in configuration $C_{\tau}$ at time $\tau$ when at time $t$ the total integrated 
current is $Q_t$. Timescales are such that $1 \ll \tau \ll t$, so all times involved are long enough for the memory of the initial state $C_0$ to be lost. We can write now
\begin{widetext}
\begin{equation}
\bar{P}(C_{\tau},Q_t,\tau,t) = \sum_{C_{t}\ldots C_{\tau+1} C_{\tau-1} \ldots  C_1} U_{C_t C_{t-1}} \cdots U_{C_{\tau+1} C_{\tau}} 
U_{C_{\tau} C_{\tau-1}} \cdots U_{C_1 C_0}\, \delta \Big(Q_t - \sum_{k=0}^{t-1}J_{C_{k+1} C_k}\Big) \, ,
\end{equation}
\end{widetext}
where we do not sum over $C_{\tau}$. Defining the moment-generating function of the above distribution, 
$\bar{\Pi}(C_{\tau},\lambda,\tau,t)=\sum_{Q_t} \exp (\lambda Q_t) \bar{P}(C_{\tau},Q_t,\tau,t)$, 
we can again check that the probability weight of configuration $C_{\tau}$ at intermediate
time $\tau$ in a large deviation event of current $q=Q_t/t$, $P_q^{\text{mid}}(C_{\tau}) \equiv \bar{P}(C_{\tau},Q_t,\tau,t)/P(Q_t,t)$, is also given by 
$P_{\lambda}^{\text{mid}}(C_{\tau})\equiv \bar{\Pi}(C_{\tau},\lambda,\tau,t)/\Pi(\lambda,t)$ for long times such that $1 \ll \tau \ll t$, with $q=q_o(\lambda)$. 
In this long-time limit one thus finds
\begin{equation}
P_{\lambda}^{\text{mid}}(C_{\tau}) \propto \la \Lambda^L(\lambda) | C_{\tau} \ra \la C_{\tau} | \Lambda^R(\lambda) \ra \, ,
\label{profmid1}
\end{equation}
in contrast to $P_{\lambda}^{\text{end}}(C)$, which is proportional to $\la C | \Lambda^R(\lambda)\ra$, see above.
Here $|\Lambda^R(\lambda)\ra$ and $\la \Lambda^L(\lambda)|$ are the right and left eigenvectors associated to the largest eigenvalue $\text{e}^{\Lambda(\lambda)}$ 
of modified transition rate $\tilde{U}(\lambda)$, respectively. They are different because $\tilde{U}$ is not symmetric. In order to compute the left eigenvector, notice that 
$|\Lambda^L(\lambda) \ra$ is the \emph{right} eigenvector of the transpose \emph{matrix} $\tilde{U}^{\text{T}}(\lambda)$ with eigenvalue 
$\text{e}^{\Lambda(\lambda)}$. 
This right eigenvector of $\tilde{U}^{\text{T}}(\lambda)$ can be in turn related to the corresponding right eigenvector of $\tilde{U}(-\lambda-{\cal E})$ by noticing that
the local detailed balance condition holds for the KMP model, guaranteeing the time reversibility of microscopic dynamics. 
This condition states that 
$U_{C' C} p_{\text{eq}}(C) = U_{C C'} p_{\text{eq}}(C')\text{e}^{{\cal E} J_{C' C}}$, where
$p_{\text{eq}}(C)$ is an effective equilibrium weight which for the KMP model takes the value $p_{\text{eq}}(C)= \exp (-\sum_{y=1}^N \beta_y e_y)$ 
with $C=\{e_y, y=1\ldots N\}$ and $\beta_y=T_L^{-1} + {\cal E} \frac{y-1}{N-1}$.
Local detailed balance then implies a symmetry between the forward modified dynamics for a current fluctuation and the time-reversed modified dynamics for the negative current fluctuation, i.e. $\tilde{U}_{C C'}=p_{\text{eq}}^{-1}(C') \tilde{U}(-\lambda-{\cal E})p_{\text{eq}}(C)$, or in matrix form
\begin{equation}
\tilde{U}^{\text{T}} (\lambda) = \mathbf{P}_{\text{eq}}^{-1} \tilde{U}(-\lambda-{\cal E}) \mathbf{P}_{\text{eq}} \, ,
\label{transpose}
\end{equation}
where $\mathbf{P}_{\text{eq}}$ is a diagonal \emph{matrix} with entries $p_{\text{eq}}(C)$.
Eq. (\ref{transpose}) implies that all eigenvalues of 
$\tilde{U}(\lambda)$ and $\tilde{U}(-\lambda-{\cal E})$ are equal, and in particular the largest, so $\mu(\lambda)=\mu(-\lambda-{\cal E})$ and 
this proves the Gallavotti-Cohen fluctuation relation. Moreover, if $|\Lambda_j^R(-\lambda-{\cal E})\ra$ is a right eigenvector of 
$\tilde{U}(-\lambda-{\cal E})$, which can be expanded as $|\Lambda_j^R(-\lambda-{\cal E})\ra = \sum_C \la C | \Lambda_j^R(-\lambda-{\cal E})\ra |C\ra$,
then
\begin{equation}
|\Lambda_j^L(\lambda)\ra = \sum_C (p_C^{\text{eq}})^{-1} \la C | \Lambda_j^R(-\lambda-{\cal E})\ra |C\ra
\label{eigenleft}
\end{equation}
is the right eigenvector of $\tilde{U}^{\text{T}}(\lambda)$ associated to the same eigenvalue. In this way, by plugging this into eq. (\ref{profmid1}) we find
\begin{equation}
P_{\lambda}^{\text{mid}}(C) \propto (p_C^{\text{eq}})^{-1} \la C | \Lambda^R(-\lambda-{\cal E}) \ra \la C | \Lambda^R(\lambda) \ra \, , \nonumber
\end{equation}
where we assumed real components for the eigenvectors associated to the largest eigenvalue. Equivalently
\begin{equation}
P_{\lambda}^{\text{mid}}(C) =A \, \frac{P_{\lambda}^{\text{end}}(C) P_{-\lambda-{\cal E}}^{\text{end}}(C)}{p_C^{\text{eq}}} \, ,
\label{profmid1}
\end{equation}
with $A$ a normalization constant. This relation implies that configurations with a significant contribution to the average profile at intermediate times
are those with an important probabilistic weight at the end of both the large deviation event and its time-reversed process. Supplementing the above relation
with a local equilibrium hypothesis, one can obtain average temperature profiles at intermediate times in terms of profile statistics at the
end of the large deviation event.

\acknowledgments

We thank B. Derrida, J.L. Lebowitz, V. Lecomte and J. Tailleur for illuminating discussions and comments. Financial support 
from Spanish project FIS2009-08451, AFOSR Grant No. AF-FA-9550-04-4-22910 and University of Granada is also acknowledged.

\end{document}